\documentclass[preprint,12pt]{elsarticle}

\usepackage{amssymb}
\usepackage{amsmath}
\journal{Advances in Applied Energy}

\usepackage{import}

\usepackage{amsmath,amssymb,amsfonts}
\usepackage{graphicx}
\usepackage{textcomp}

\usepackage[dvipsnames]{xcolor}
\usepackage{bbding}
\usepackage{bm}
\usepackage{algorithm}
\usepackage{algpseudocode}
\usepackage{enumitem}
\usepackage{cases}

\newcounter{defcounter}
\newenvironment{pequation}{%
\addtocounter{equation}{-1}
\refstepcounter{defcounter}

\begin{equation}}
{\end{equation}}

\usepackage{xurl}
\usepackage{ragged2e}
\usepackage{booktabs, makecell, tabularx}

\newcolumntype{L}{>{\RaggedRight}X}
\usepackage{siunitx}
\usepackage{enumitem}

\newcolumntype{Y}{>{\centering\arraybackslash}X}
\usepackage{footnote}

\usepackage{lscape}

\usepackage{mathtools}

\newtheorem{remark}{\textbf{Remark}}

\newcommand{\rG}{\mathrm{G}}

\newcommand{\sN}{\mathcal{N}}

\newcommand{\sA}{\mathcal{A}}
\newcommand{\sG}{\mathcal{G}}

\DeclareMathOperator{\suchthat}{s.t.}

\usepackage{subcaption}

\usepackage{longtable}
\usepackage{footnote}
\makesavenoteenv{longtable}  % Ensures longtable environment works with footnotes
\usepackage{adjustbox}

\begin{document}

\begin{frontmatter}



\title{A Review of Scalable and Privacy-Preserving Multi-Agent Frameworks for Distributed Energy Resources} 
% \vspace{5cm}
% \revsecond{{\bf{REV 2.3}}}

% \author[label1,label2]{Xiang Huo}
% \affiliation[label1]{organization={},

% \affiliation[label2]{organization={},

% \author{Xiang Huo} %% Author name

\author[label1]{Xiang Huo}
\author[label2]{Hao Huang}
\author[label1]{Katherine R. Davis}
\author[label2]{H. Vincent Poor}
\author[label3]{Mingxi Liu}

% Author affiliation
\affiliation[label1]{organization={Department of Electrical and Computer Engineering, Texas A$\&$M University},%Department and Organization
            city={College Station},
            postcode={77843}, 
            state={TX},
            country={USA}}

\affiliation[label2]{organization={Department of Electrical and Computer Engineering, Princeton University},%Department and Organization
            city={Princeton},
            postcode={08544}, 
            state={NJ},
            country={USA}}

\affiliation[label3]{organization={Department of Electrical and Computer Engineering, University of Utah},%Department and Organization
            city={Salt Lake City},
            postcode={84112}, 
            state={UT},
            country={USA}}

%% Abstract

\begin{abstract}
Distributed energy resources (DERs) are gaining prominence due to their advantages in improving energy efficiency, reducing carbon emissions, and enhancing grid resilience. Despite the increasing deployment, the potential of DERs  has yet to be fully explored and exploited. A fundamental question restrains the management of numerous DERs in large-scale power systems, ``\textit{How should DER data be securely processed and DER operations be efficiently optimized?}'' To address this question, this paper considers two critical issues, namely \textit{privacy} for \textit{processing DER data} and \textit{scalability} in \textit{optimizing DER operations}, then surveys existing and emerging solutions from a multi-agent framework perspective. In the context of scalability, this paper reviews state-of-the-art research that relies on parallel control, optimization, and learning within distributed and/or decentralized information exchange structures, while in the context of privacy, it identifies privacy preservation measures that can be synthesized into the aforementioned scalable structures. Despite research advances in these areas, challenges remain because these highly interdisciplinary studies blend a wide variety of scalable computing architectures and privacy preservation techniques from different fields, making them difficult to adapt in practice.  To mitigate this issue, this paper provides a holistic review of trending strategies that orchestrate privacy and scalability for large-scale power system operations from a multi-agent perspective, particularly for DER control problems. Furthermore, this review extrapolates new approaches for future scalable,  privacy-aware, and cybersecure pathways to unlock the full potential of DERs through controlling, optimizing, and learning generic multi-agent-based cyber-physical systems.
\end{abstract}


% \begin{highlights}

% \item We conduct and present a systematic review of deploying multi-agent frameworks for distributed energy resource (DER) control in power systems regarding \textit{multi-agent-based problem formulation}, \textit{scalable solutions}, and \textit{privacy preservation techniques}.

% \item  We survey state-of-art scalable algorithms within multi-agent frameworks based on distributed and decentralized information exchange structures, and we review representative works for DER control problems. Moreover, we identify internal, external, and hierarchical types of adversaries/threats in multi-agent systems that can compromise the system's privacy and security.

% \item We categorize representative \textit{privacy preservation techniques} into \textit{differential privacy}, \textit{cyptographic methods}, and other \textit{miscellaneous and emerging methods}, and discuss their features and applications to adapt into the scalable and privacy-preserving DER control.

% \item Building on the summarization and discussion of existing works, this review extrapolates new approaches for future scalable, privacy-aware, and cybersecure multi-agent frameworks to unlock the full potential of DERs. These directions include \textit{enhancing accuracy, privacy, and algorithm efficiency}, \textit{establishing trustworthiness across fields}, and \textit{developing zero-trust standards}.

% \end{highlights}

%% Keywords

\end{frontmatter}

\section{Introduction}
\label{introduction}
% \newpage

\subsection{Significance of DERs}

Distributed energy resources (DERs), including solar photovoltaics (PVs), wind turbines, fuel cells, energy storage systems (ESSs), and electric vehicles (EVs), refer to a variety of small-scale energy generation and storage devices that are connected to the electric power grid \cite{dernerl}. 
% \revsecond{{\bf{REV 2.1}}}{\color{red}
% 
DERs offer substantial flexibility to power systems at both the grid and customer levels, such as providing ancillary services \cite{wang2015review}, lowering energy costs \cite{lee2021autoshare}, decarbonizing power systems \cite{denholm2022examining,akorede2010distributed}, and enhancing grid resilience \cite{basak2012literature}. 
% %
Because of these benefits, the power grid is transitioning toward an increasingly DER-rich electricity system, where the management of DERs is crucial for supporting the integration of renewable energy, enhancing grid resilience, and improving overall energy efficiency \cite{highDERinter}. 
As the energy transition accelerates, the growing dependence on DERs is reshaping how loads and generation sources are managed for homes and utilities, offering cheaper, cleaner, and more accessible electricity supplies. 
Additionally, past grid failures from extreme weather events, operational breakdowns, and cyber-physical attacks \cite{panteli2015influence,shi2022enhancing,abdelmalak2022enhancing,habib2017review,gouveia2009evaluating,deng2016false,lai2017cyber} have underscored the need for system operators (SOs), prosumers, and consumers to increasingly rely on DERs, both individually and in aggregate, to bolster grid resilience.
% 
% }
% 
% 
The global DER management system market is expected to expand significantly, projected to increase from USD 0.42 billion in 2021 to USD 1.33 billion by 2028, reflecting a compound annual growth rate of 18.0\% during this period \cite{Fortune_Business_Insights_DER}. 
In the U.S., the DER market is anticipated to nearly double in capacity from 2022 to 2027, with capital expenditures reaching USD 68 billion per year \cite{woodmacDER2023}. 
% 
% \hspace{0mm}\revsecond{{\bf{REV 2.1}}}{\color{red}
%
Hence, the importance of DERs in power systems can be seen by their rapid growth and their pivotal role in modernizing the grid. 
% 
% }
% The power grid is transitioning towards a DER-rich electricity system, where the management of DERs is crucial for supporting the integration of renewable energy, enhancing grid resilience, and improving overall energy efficiency \cite{highDERinter}. 

% in achieving grid sustainability, resilience, and cybersecurity 

% {\color{red}
\subsection{Major DER Control Challenges}
% }
% \deleted{Advanced control, optimization, and machine learning theories and tools are essential to fully realize the potential of DERs, especially for achieving scalability in large-scale power systems. These methodologies can assist in solving the DER management problem via generic mathematical formulations with grid objectives and constraints. Broadly, the scalable control of DERs within power systems can be interpreted through a networked multi-agent (we refer to an element of a DER system as an {\it agent}) problem where agents can operate in parallel. The unprecedented
% deployment rates of DERs require 
% scalable management solutions on grid-tied resources to achieve full decarbonization at scale \cite{denholm2022examining}. 
% % 
% Besides,  building
% scalable deployment models can accelerate the adoption of the key commercially available but underutilized grid solutions
% needed to maintain a reliable, safe, and affordable grid \cite{DOE_grid_liftoff}.}

% {\color{red}
\subsubsection{Scalability Issues}
% \hspace{0mm}\revsecond{{\bf{REV 2.2}}}
% 
Traditionally, the power grid is managed in a centralized manner, with a single control problem formulated using data collected from the wide-area power grid network.
However, the rapid increase of DERs at various scales makes it challenging to solve such large-scale centralized problems, due to the growing complexity, amount of heterogeneous data, and high computing costs. 
The high deployment of DERs requires scalable management solutions for grid-tied resources to meet grid objectives, such as full decarbonization at scale \cite{denholm2022examining}. 
Besides, building scalable deployment models can accelerate the adoption of commercially available grid solutions  \cite{DOE_grid_liftoff}.
% }
% 
Broadly, the scalable control of DERs within power systems can be interpreted through a networked multi-agent (we refer to an element of a DER system as an {\it agent}) problem where agents can operate in parallel. 
% 
% \revsecond{{\bf{REV 2.2}}}{\color{red}
To fully realize the potential of DERs to help achieve greater sustainability and resilience in
power systems, it is essential to develop scalable multi-agent frameworks incorporating advanced control, optimization, and machine learning theories and tools. 
These advanced techniques can help solve the DER management problems efficiently via generic mathematical formulations in a scalable fashion that incorporates grid objectives and constraints.
For example, in \cite{radhakrishnan2016multi}, Radhakrishnan and Srinivasan propose a multi-agent-based DER management system to perform optimal management of DERs. Wang \textit{et al.} \cite{wang2020data} propose a data-driven multi-agent power grid control scheme using a deep reinforcement learning (RL) method for autonomous voltage control. In \cite{ fan2023multi}, Fan \textit{et al.} develop a multi-agent distributed control framework to realize the distributed optimal generation control with multiple DERs in real-time. All these developments highlight the benefits of developing multi-agent frameworks with advanced techniques for DER management to ensure the optimality, safety, and security of power grid operations.
% }

% {\color{red}
\subsubsection{Privacy Threats}
% }
% 
Another key consideration is privacy preservation/protection. Privacy breaches can happen during the processing and transmission of DER data, such as malicious interception of private information during data transmission and the loss of data provenance in the face of dishonest agents \cite{lu2018privacy, li2019privacy, zhang2022privacy, wang2022differentially}. 
By analyzing load data, adversaries can infer consumers' lifestyle patterns, personal preferences, and occupancy profiles \cite{lisovich2010inferring,greveler2012multimedia,chin2017privacy}, thereby leading to privacy risks. 
The threat of privacy leakages targeting the power electric sector, especially DER-rich electric power grids, is escalating in both frequency and complexity. 
In recent years, a series of stringent privacy protection laws have come into effect to  increase protections for consumers' personal data. These include the strongest privacy and security law  \cite{GDPR}, \emph{European Union’s General Data Protection Regulation}, effective in 2018, the U.S.'s first privacy law \emph{California Consumer Privacy Act} \cite{ccpa_california}, also effective in 2018, the Virginia's \emph{Consumer Data Protection Act} \cite{cdpa_virginia}, effective in 2023, and the most recent \emph{Texas Data Privacy and Security Act} \cite{TDPSA_texas}, effective in 2024, and other similar laws \cite{more_privacy_laws}. The increased privacy awareness in legislation is driving privacy protection measures and standards in a broad spectrum of cyber and physical systems, such as health care systems \cite{sahi2017privacy}, wireless networks \cite{li2009privacy}, and power systems in this regard \cite{erkin2013privacy}.  
% 
% \revfirst{{\bf{REV 1.1}}}{\color{blue}
% 
For the cybersecurity of smart grids, National Institute of Standards \& Technology (NIST) has established the \emph{Guidelines for Smart Grid Cybersecurity} to develop effective cybersecurity strategies for protecting the privacy of smart grid-related data and for securing the communication networks \cite{NIST_SmartGrid}. The U.S. Department of Energy, in collaboration with industry stakeholders, launched the \emph{DataGuard Energy Data Privacy Program} to develop privacy concepts and principles for grid customers \cite{DOE_dataguard}. Europe Commission has introduced the \emph{Data Protection Impact Assessment} template for smart grid and smart metering systems, aimed at assessing risks and developing countermeasures to ensure the protection of personal data \cite{EU_dpia}. Additionally, the International Electrotechnical Commission (IEC) has specified cybersecurity requirements for smart grids through a series of standards in \emph{IEC 62351}, addressing secure data transfer, prevention of eavesdropping, and intrusion detection \cite{iec_62351_standards}.
Therefore, eliminating privacy and security concerns is critical for deploying advanced multi-agent frameworks to optimize DER operations.  
% }

% {\color{red}
\subsection{Motivation and Contributions}
% }
% {\color{red}
To unlock the full potential of DERs, this paper focuses on two key technical challenges in optimizing DER-rich power systems, i.e., scalability and privacy. 
% }
% \deleted{
The old model of centralized electrical supply is no longer the sole reality, massive DERs with varying attributes require a scalable management plan. Furthermore, to capture the DER market and enhance customer engagement, privacy-preserving decision-support tools need to be developed, integrated, and tested. These tools, in turn, can optimize customers' electrical energy services. The synergy of scalability and privacy protection is becoming a trending research topic among the control, optimization, learning, and power communities.
% } 
% 
A number of existing reviews have underlined the importance of DER control, along with arising related \textit{privacy} and \textit{cybersecurity} concerns \cite{zografopoulos2023distributed,ghiasi2023comprehensive,ferrag2018systematic,tuyen2022comprehensive,hasan2023review,liu2024enhancing,sebastian2023review}. 
Zografopoulos \emph{et al.} \cite{zografopoulos2023distributed} 
point out cybersecurity issues in DER control problems caused by adversarial capabilities and objectives, and review both DER protocol-level and DER device-level vulnerabilities, attacks, impacts, and mitigations.
Ghiasi \emph{et al.} \cite{ghiasi2023comprehensive} provide a comprehensive review of cyberattacks and defense mechanisms for smart grid energy systems. In \cite{ferrag2018systematic}, the authors review privacy-preserving schemes for smart grid applications, addressing privacy leakages from key-based, data-based, impersonation-based, and physical-based attacks. Considering the increasing penetration of renewable energy, Tuyen \emph{et al.} \cite{tuyen2022comprehensive} survey state-of-the-art detection and mitigation techniques, with a focus on the system structure and vulnerabilities of typical inverter-based power systems integrated with DERs. 
In \cite{hasan2023review}, the authors review standards, protocols, and constraints and provide recommendations for mitigating cyber-attacks in cyber-physical power systems. 
Liu \emph{et al.} \cite{liu2024enhancing} survey developments on enhancing cyber-resiliency of DER-based smart grids, including threat modeling, risk assessment, and defense-in-depth strategies. In \cite{sebastian2023review}, Cardenas \emph{et al.}  underscore the need for privacy-aware solutions and discuss grid-related digital privacy risks. 
Notably, the aforementioned reviews provide different examinations of controlling DERs from multi-faceted perspectives. However, an interdisciplinary review of orchestrated scalable and privacy-preserving solutions is important to deploy advanced multi-agent DER control strategies in practice.

Motivated by the proliferation of recent research outcomes on DER controls, this paper reviews state-of-the-art techniques for designing scalable and privacy-preserving multi-agent frameworks and their applications to DER control problems. 
\begin{figure*}[ht]
 \centering\includegraphics[width=1\textwidth]{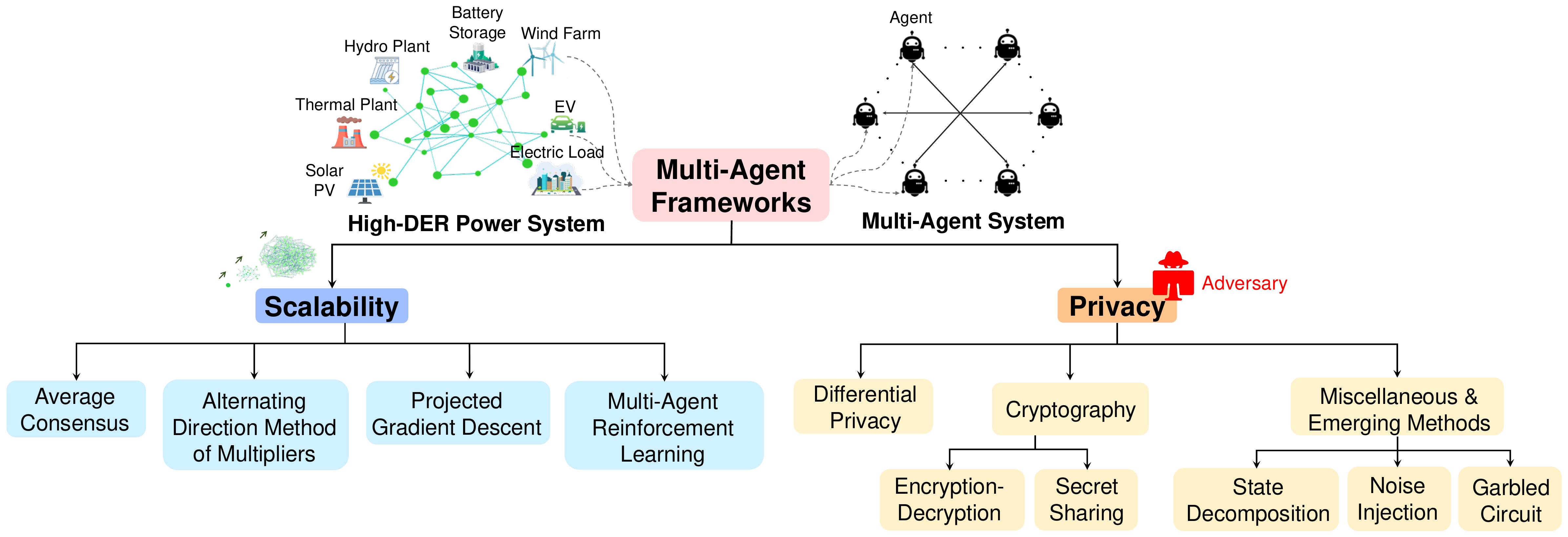}
\caption{Review structure of \textit{scalable} and \textit{privacy-preserving} multi-agent frameworks for DERs.}
% \revsecond{{\bf{REV 2.5}}}
\label{fig_paper_structure}
\end{figure*}  
Fig.~\ref{fig_paper_structure} provides an overview of the review structure. We first survey scalable multi-agent frameworks based on contemporary distributed and decentralized information exchange structures, and then review integrated privacy preservation techniques from the perspective of privacy-aware multi-agent computing frameworks. 
%

% \begin{figure}[htbp]
%  \centering\includegraphics[width=0.48\textwidth]{figures/Privacy_issues_architecture.eps}
% \caption{Overview of the review structure on \textit{scalability} and \textit{privacy} for multi-agent frameworks.}
% \label{fig_paper_structure}
% \end{figure} 
%
To the best of our knowledge, this paper, for the first time, surveys the effectiveness of scalability and privacy preservation ability in  distributed and decentralized multi-agent frameworks, with an emphasis on large-scale DER control applications.  The contributions of this paper include:
\begin{enumerate}

\item We conduct and present a systematic review of deploying multi-agent frameworks for DER control in power systems regarding \textit{multi-agent-based problem formulation}, \textit{scalable solutions}, and \textit{privacy preservation techniques}.

\item  We survey state-of-art scalable algorithms within multi-agent frameworks based on distributed and decentralized information exchange structures, and we review representative works for DER control problems. Moreover, we identify internal, external, and hierarchical types of adversaries/threats in multi-agent systems that can compromise the system's privacy and security.

\item We categorize representative \textit{privacy preservation techniques} into \textit{differential privacy}, \textit{cyptographic methods}, and other \textit{miscellaneous and emerging methods}, and discuss their features and applications to adapt into the scalable and privacy-preserving DER control.

\item Building on the summarization and discussion of existing works, this review extrapolates new approaches for future scalable, privacy-aware, and cybersecure multi-agent frameworks to unlock the full potential of DERs. These directions include \textit{enhancing accuracy, privacy, and algorithm efficiency}, \textit{establishing trustworthiness across fields}, and \textit{developing zero-trust standards}.

\end{enumerate}

% \hspace{0mm}\revsecond{{\bf{REV 2.4}}}{\color{red}
% 
In the rest of this paper, Section \ref{Preliminaries_on_DER_Control} provides an overview of multi-agent systems for DER control applications in power systems, detailing the corresponding multi-agent optimization model and the information exchange structures. 
% }
% \replaced{Section \ref{Preliminaries_on_DER_Control} }{\ref{background}} gives an overview of multi-agent systems for DER control applications in power systems, with details of constructing a multi-agent optimization model and information exchange structure. 
% 
Section \ref{Scalable_Methods} surveys predominantly scalable methods for addressing the multi-agent problem and summarizes privacy issues associated with these scalable approaches. 
Section \ref{Privacy_Preservation_Techniques} reviews privacy preservation techniques for scalable multi-agent-based DER control. 
Section \ref{Future_Directions} extrapolates new approaches for future scalable, privacy-aware, and cybersecure pathways to unlock the full potential of DERs. 
Section \ref{Conclusion} concludes the paper. 

\section{Multi-Agent Systems of DERs}
\label{Preliminaries_on_DER_Control}
% }

The DER control problem can be described using a multi-agent optimization model that defines the objectives, constraints, and decision variables to optimize the power grid operations. The optimal decision-making can be achieved by solving the formulated  multi-agent problem with control, optimization, and learning-based methods. In this section, we present a general multi-agent problem formulation and then delve into the detailed objectives and constraints for multi-agent-based DER control problems. 

% {\color{red}
\subsection{Multi-Agent-Based Control of DERs}
% \label{background}
% \revsecond{{\bf{REV 2.4}}}} 
% }
%\subsection{Multi-agent Frameworks for DERs Operations in Power Grids}
%\subsection{Background of DER}

%\subsection{Background}
%To fully unleash the potential of DERs, advanced control, optimization, and learning theories and tools are required, especially for managing DERs in large-scale power systems. These methodologies can assist in solving the DER management problem via mathematical formulations with grid-oriented objectives and constraints. Generally speaking, 

The management of DERs in power systems can be viewed as the control of agents within a networked multi-agent system. To describe such a multi-agent system, we need to define an \emph{Optimization model} that specifies the problem objectives and constraints and an \emph{Information exchange model} that details the agents’ information exchange structure \cite{nedic2009distributed}. 
The optimization model includes cooperative (for the system) and/or competitive objectives (between agents) and is subject to networked constraints (related to a set of agents) and/or local constraints (related to only an individual agent). For the DER control problems in power systems, we classify the objectives into two categories, including \emph{cooperative grid-level objective} and \emph{competitive DER-level objective}.

The cooperative grid-level objectives support the achievement of system-wide goals, such as achieving overnight valley filling \cite{gan2012optimal,zhang2014coordinating,jian2017high}, minimizing power lines losses \cite{sultana2016review,rao2012power,kashem2000novel}, reducing the emission of pollutants \cite{denny2006wind,khan2018importance}, etc.  The competitive DER-level objectives aim at maximizing the benefits of DERs, such as bidding in the electricity market \cite{xiao2020coordination}, reducing energy costs for consumers \cite{akhter2023efficient,hu2023economic}, and minimizing battery degradation costs \cite{chen2023optimal}. 
The networked constraints can include the nodal voltage deviations and  power flow constraints \cite{dommel1968optimal,Shining2024gridresponder}, which are coupled through the power network topology. The individual DER constraints can include battery charging/discharging rate \cite{awad2014optimal}, the capacity of generators \cite{wu2016distributed}, solar power availability for PV curtailment \cite{wang2015dynamic}, etc. 
% \revfirst{{\bf{REV 1.2}}}{\color{blue}
Besides, individual DER constraints often need to account for thermal dynamics, which are coupled with the operation of heating, ventilation, and air conditioning (HVAC) systems \cite{li2011optimal,huo2022two}.
% }

% \begin{figure*}[!htb]
% % \vspace*{-2mm}
%     \centering
%     \includegraphics[width=0.85\textwidth,trim = 0mm 0mm 0mm 0mm, clip]{figures/central_distri_decentr.eps}
%     % \vspace*{-3mm}
%     \caption{Three typical information exchange structures for networked multi-agent systems to manage DERs in power systems:  (a) \emph{Centralized} information exchange that relies on a system operator to collect information from all agents, process it, and then send control commands to each agent; (b) \emph{Distributed} structure that allows agents to operate independently, interact with coordinator/environment, and communicate with each other over a network; and (c) \emph{Decentralized} structures that is similar to a \emph{Distributed} structure, but without peer-to-peer communications. \textcolor{red}{add (a), (b), (c) in the figure?}}
%     % \vspace*{-4mm}
%     \label{central_distri_decentr}
% \end{figure*}

\begin{figure*}[!htb] 
\centering
\begin{subfigure}[t]{0.3\textwidth}
\centering
\includegraphics[width=0.85\textwidth,trim = 0mm 0mm 0mm 0mm, clip]{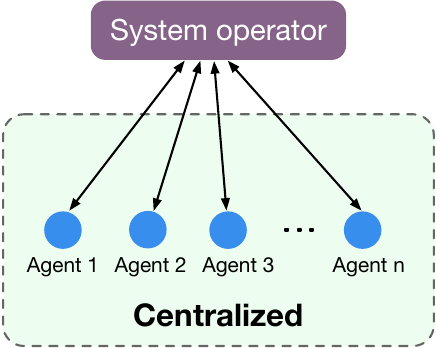}
\caption{Centralized structure.}
\end{subfigure}%
~ 
\begin{subfigure}[t]{0.3\textwidth}
\centering
\includegraphics[width=0.85\textwidth,trim = 0mm 0mm 0mm 0mm, clip]{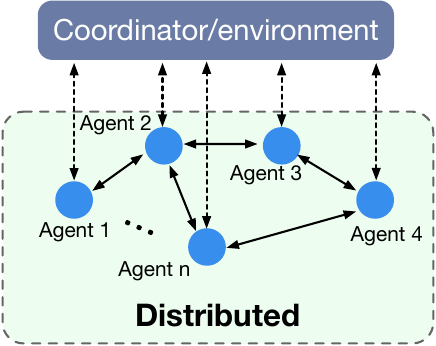}
\caption{Distributed structure.}
\end{subfigure}
~ 
\begin{subfigure}[t]{0.3\textwidth}
\centering
\includegraphics[width=0.85\textwidth,trim = 0mm 0mm 0mm 0mm, clip]{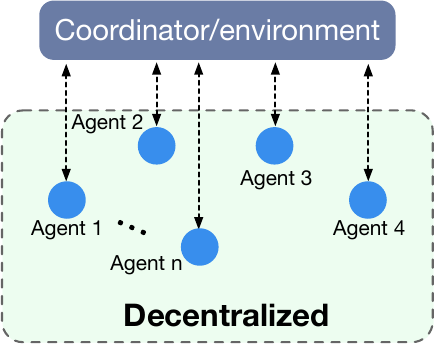}
\caption{Decentralized structure.}
\end{subfigure}
\caption{Three typical information exchange structures of networked multi-agent systems for managing DERs in power systems:  (a) \emph{Centralized} information exchange that relies on a system operator to collect information from all agents, process it, and then send control commands to each agent; (b) \emph{Distributed} structure that allows agents to operate independently, interact with coordinator/environment, and communicate with each other over a network; and (c) \emph{Decentralized} structures that is similar to a \emph{Distributed} structure, but without agent-to-agent communications.}
\label{central_distri_decentr}
\end{figure*}

The information exchange model defines the computing and communication structure for solving networked multi-agent problems.  Various models have been developed to facilitate  control, optimization, and learning in these  multi-agent systems \cite{jian2017high,boyd2011distributed,colfati2007consensus,bertsekas2015parallel,koshal2011multiuser,liu2013decentralized,zhang2021distributed}. In summary, these models can be classified into \emph{centralized}, \emph{distributed}, and \emph{decentralized} structures, as shown in Fig.~\ref{central_distri_decentr}. 
% \revfirst{{\bf{REV 1.3}}}{\color{blue}
Multi-agent frameworks also include \emph{hierarchical} information exchange structures that combine centralized and distributed computing schemes, leveraging the strengths of both approaches to achieve both global oversight and local autonomy. 
In a hierarchical framework, a centralized layer typically oversees high-level decision-making and coordination by utilizing global optimization to guide agents toward optimal solutions.
On the other hand, the distributed layer allows individual agents to operate autonomously, handling localized tasks and reacting to dynamic environmental changes in real-time.
% }
Following this classification, this paper reviews scalable multi-agent frameworks within distributed and decentralized structures, addressing the interests of different stakeholders when operating DER-rich power systems. The system operator (SO) (e.g., distribution or transmission SO) functions as a central authority and can provide instructions on coordinating and controlling the power system operations. To clarify, we refer to the coordinator as a central entity that is needed solely for the coordination of signals rather than for directly controlling any agent.

In a centralized setting, the SO manages the entire power system operation by collecting agent and network information, processing it, and sending control commands to all agents \cite{jian2017high}. Therefore, the DER control problem is solved centrally where the SO makes strategic decisions on achieving grid-level and/or DER-level objectives, while agents simply follow the SO's commands. Centralized approaches are easy to implement and can often obtain globally optimized solutions. However, they suffer from drawbacks caused by 1) computing and communication overheads imposed on the SO, 2) compromised data privacy and security, and 3) vulnerability during cyber and physical contingencies. Due to the heavy dependence on the SO, centralized information exchange is more effective for small-scale cases with a modest agent population size.

In contrast to centralized methods, distributed and decentralized information exchange structures offer scalability, resilience, and enhanced privacy and cybersecurity, especially when applied to power systems with large DER populations, complex network topologies, and sophisticated control procedures   \cite{boyd2011distributed,colfati2007consensus,bertsekas2015parallel}. In a distributed setting, the original large-scale problem is decomposed into small-scale subproblems where each agent exchanges information with other agents (e.g., its adjacent neighbors) to update its decision variables. Parallel computing is implemented at local agents, such as through the alternating direction method of multipliers (ADMM) \cite{boyd2011distributed}, thereby offering high scalability for solving large-scale DER control problems. By enabling efficient information exchange and parallel local computing among agents, distributed structures eliminate the need for agents to rely solely on the SO when making decisions.

Similarly, decentralized approaches also achieve scalability by allocating central computing loads to each local agent, but with an emphasis on eliminating agent-to-agent communications. In decentralized information exchange structures, agents make decisions independently in a possible networked environment without communicating with each other. 
% \revfirst{{\bf{REV 1.4}}}{\color{blue} 
Compared to distributed structures, decentralized approaches eliminate agent-to-agent communication, thereby reducing privacy risks associated with direct information exchange between agents.
However, agents in decentralized structures are often required to interact directly either with the environment or rely on the assistance of a coordinator, both of which can lead to private information leakage, such as in the presence of \emph{honest-but-curious agents}, \emph{external eavesdroppers}, and/or \emph{the coordinator}.
% }  
In a typical decentralized framework, such as primal-dual-based algorithms \cite{koshal2011multiuser,liu2013decentralized}, agents and the coordinator iteratively update the primal variable (decision variable) and the dual variable (Lagrange multiplier), respectively. Owing to the outstanding scalability, distributed and decentralized multi-agent frameworks are well suited for large-scale DER control problems.

As shown in Fig. \ref{central_distri_decentr}, the frequent and mandated exchange of private information in centralized, distributed, and decentralized multi-agent frameworks renders the system and agents vulnerable to privacy breaches. 
The acquisition, processing, and transmission of private customer data are typically necessary for delivering grid services and enhancing customer satisfaction \cite{Gough9411689,dvorkin2020differentially,dong2018privacy,atmaca2024privacy}.
However, unauthorized processing and sharing of sensitive information can result in privacy leakages and malicious manipulation of the system, introducing vulnerabilities that hinder the deployment of advanced DER control approaches.
To protect the privacy of stakeholders, it is essential to integrate privacy preservation techniques into the design of scalable multi-agent frameworks. To this end, we identify typical adversaries in distributed and decentralized multi-agent frameworks from internal, external, and hierarchical perspectives. These adversaries present distinct threats with varying attack vectors \cite{huang2012differentially,nozari2016differentially,lu2018privacy,hadjicostis2020privacy,wu2021privacy,li2019privacy,huo2022distributed,tian2022secret,wang2019privacy,ben2008fairplaymp,charalambous2019privacy}, including \emph{Honest-but-curious agents} who do not interfere with the algorithm but may use the accessible information to infer the private data of other participants, \emph{External eavesdroppers} who wiretap the exchanged messages between agents and/or the SO/coordinator/aggregator, and \emph{the SO/coordinator/aggregator} who directly communicates with and/or controls the agents and has their private data. 
Consequently, resolving privacy challenges in multi-agent systems has become a burgeoning research topic, driving the development of privacy-preserving frameworks that ensure privacy guarantees across diverse operational scenarios for DER control problems.

% {\color{red}
\subsection{General Problem Formulation}

\begin{figure}[!htb]
% \vspace*{-2mm}
    \centering
    \includegraphics[width=0.8\textwidth, trim = 2mm 0mm 1.5mm 0mm, clip]{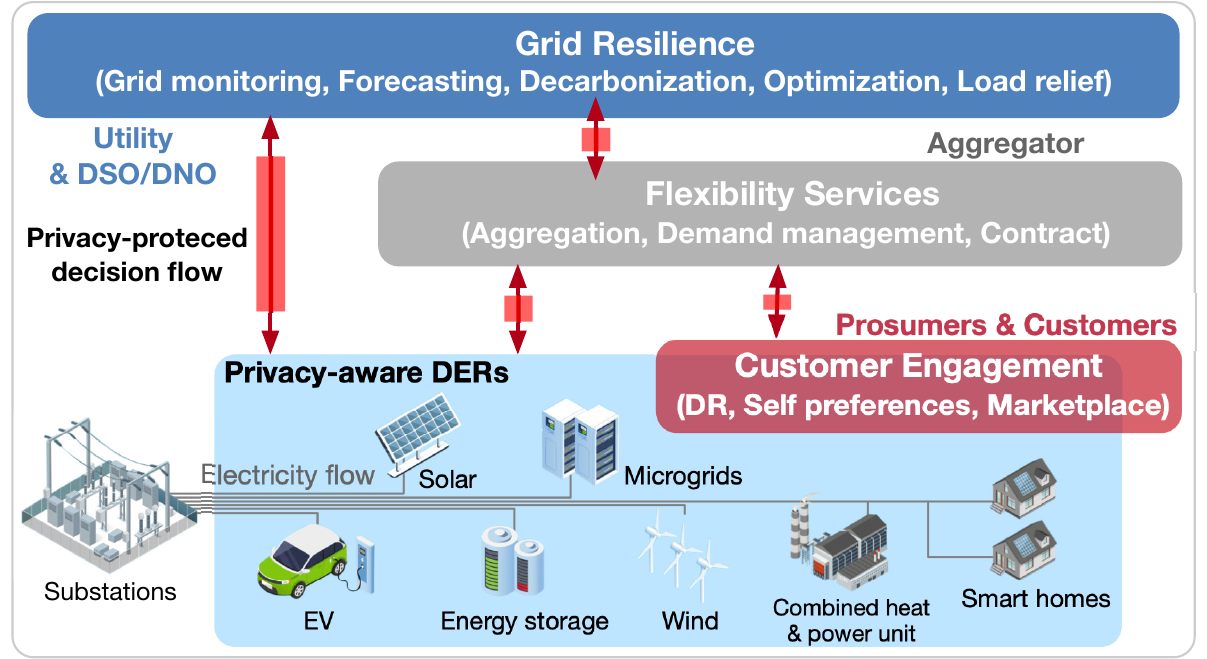}
    % \vspace*{-3mm}
    \caption{Conceptualization of a privacy-preserving DER management system.}
    % \vspace*{-4mm}
    \label{privacy_DERMS}
\end{figure}

Generically, the DER control problem (e.g., DER management system elicited by Fig. \ref{privacy_DERMS}) can be framed into a multi-agent setting, with decision variables (e.g., charging/discharging of batteries and flexible loads), cooperative (grid-level) and competitive (DER-level) objectives, network models (e.g., power distribution and transmission networks), network constraints (e.g., current and voltage constraints), and individual constraints (e.g., DER's operational constraints). Fundamentally, we provide a generic optimization model that can describe the DER control problem as: 
\begin{subequations} \label{DER_problem}
\begin{align}
 \underset{\textsc{Decision Variables}}{\text{Optimize}} \ \ \  & \text{Cooperative} + \text{Competitive}   \label{eq:copf:obj}\\
 \suchthat \ \; \ \ \ \ 
 & \text{Network Models}\\
& \text{Network Constraints} \\
& \text{Individual Constraints}
\end{align}
\end{subequations}

Problem \eqref{DER_problem} can be broadly applied to a variety of power system applications, with goals such as grid modernization, decarbonization, and resilience. 
Fig. \ref{privacy_DERMS} shows the conceptualization of a privacy-preserving DER management system, where the prosumers, customers, aggregators, utilities, and distribution network/system operators (DNO/DSO) collaborate to achieve grid-level objectives and customer-side goals. 
In the following section, we provide details on multi-agent-based DER control using the formulation of Problem \eqref{DER_problem}.

\subsection{Network Models, Objectives, and Constraints}

\subsubsection{Network Models}

\paragraph{Power flow model} 
The power flow model is built upon power network topology, loads, and generation sources. DERs can function as flexible loads or generation units in power distribution and transmission networks.
We next present the control of DERs in distribution systems using the nonlinear DistFlow branch model \cite{baran1989network}. 
Consider a radial distribution network described by a connected graph $G=\{\mathcal{N},\mathcal{E}\}$,
where $\mathcal{N} = \{0,1,\ldots,n\}$ denotes a set of nodes/buses, $\mathcal{E} \subset \mathcal{N} \times \mathcal{N}$ denotes a set of directed edges/lines. The network is tree-structured where Node 0 serves as the slack bus and $V_0$ denotes the constant voltage magnitude of Node 0.

Let $V_j$ denote the voltage magnitude of Node $j$,  $\mathcal{C}_j$ denote the set of children of Node $j$, and let the line $l_{jk} \in \mathcal{E}$ connect two neighboring nodes, Node $j$ and Node $k$. The active and reactive power flows from Node $i$ and Node $j$ are represented by $\mathcal{P}_{ij}$ and $\mathcal{Q}_{ij}$, respectively, the resistance and reactance of the line $l_{ij}$ are given by $r_{ij}$ and $x_{ij}$, respectively. Let $P_i$, $Q_i$, $p_i$, and $q_i$ denote the active power consumption, reactive power consumption, active power injection, and reactive power injection to Node $i$, respectively.

The DistFlow branch equations can be written in the real form as \cite{baran1989network,farivar2013equilibrium}:
% \begin{subequations}
% \begin{align}
%     &\mathcal{P}_{ij} - \sum_{k \in \mathcal{C}_j} \mathcal{P}_{jk} = P_j  - p_j + r_{ij}\mathcal{I}_{ij}^2, &&\forall i\in \mathcal{E} \label{distflow_1}  \\
%     &\mathcal{Q}_{ij} - \sum_{k \in \mathcal{C}_j} \mathcal{Q}_{jk} = Q_j  - q_j + x_{ij}\mathcal{I}_{ij}^2, &&\forall i\in \mathcal{E} \label{distflow_2}  \\
%     &V_i^2 - V_j^2 = 2(r_{ij} \mathcal{P}_{ij} + x_{ij}\mathcal{Q}_{ij}) \nonumber\\
%     & \qquad\qquad\qquad -(r_{ij}^2 + x_{ij}^2)\mathcal{I}_{ij}^2, &&\forall i\in \mathcal{E} \label{distflow_3} 
% \end{align}
% \end{subequations}
\begin{subequations}
\label{distflow} 
\begin{align}
     \sum_{k \in \mathcal{C}_j} \mathcal{P}_{jk} &= \mathcal{P}_{ij} - P_j  + p_j - r_{ij}\mathcal{I}_{ij}^2, &&\forall j\in \mathcal{N} \label{distflow_1}  \\
     \sum_{k \in \mathcal{C}_j} \mathcal{Q}_{jk} &= \mathcal{Q}_{ij} - Q_j  + q_j - x_{ij}\mathcal{I}_{ij}^2, &&\forall j\in \mathcal{N} \label{distflow_2}  \\
    V_i^2 - V_j^2 &= 2(r_{ij} \mathcal{P}_{ij} + x_{ij}\mathcal{Q}_{ij}) \nonumber\\
    & \qquad\qquad -(r_{ij}^2 + x_{ij}^2)\mathcal{I}_{ij}^2, &&\forall ij \in \mathcal{E} \\
    \mathcal{I}_{ij}^2 &= (\mathcal{P}_{ij}^2+\mathcal{Q}_{ij}^2)/V_i^2, &&\forall ij \in \mathcal{E}
    \label{distflow_3} 
\end{align}
\end{subequations}
where $\mathcal{I}_{ij}$ denotes the current flow from Node $i$ to Node $j$.
A typical power flow problem aims to solve \eqref{distflow} for voltages and power flows, given the active and reactive power injections and line resistances and reactances.  The nonlinear DistFlow branch model can be further linearized using LinDistFlow by the approximation of $\mathcal{I}_{ij}^2 \approx 0$, given the fact that line losses are small compared to the line flows \cite{baran1989optimal}.

\paragraph{Carbon flow model} Another recently developed network model in power systems is the carbon flow model \cite{chen2023carbonfreeelectricityflowbasedframework}. To decarbonize the electric power sector, efforts have been made in carbon accounting, carbon-aware decision-making, and carbon-electricity market design \cite{kang2015carbon,chen2023carbonawareoptimalpowerflow,chen2023carbonfreeelectricityflowbasedframework}. Chen \emph{et al.} in \cite{chen2023carbonfreeelectricityflowbasedframework} introduce a flow-based emission model that is analogous to the power flows. The carbon flow model tracks carbon emissions from generators as they are transmitted through power grids, creating a virtual carbon flow within the power network. 

Specifically, the carbon flow model is defined via the concept of nodal carbon intensity as \cite{chen2023carbonawareoptimalpowerflow}:
\begin{align}\label{eq:mci}
  w_i  \!= \! \frac{R_i^{\mathrm{in}}}{P_i^{\mathrm{in}}}\!=\!  \frac{ \sum_{g\in\sG_i} w_{i,g}p_{i,g}^{\rG} + \sum_{k\in \sN_i^+ } w_k P_{ki}}{\sum_{g\in\sG_i}  p_{i,g}^{\rG} + \sum_{k\in {\sN}_i^{+} } P_{ki} },\ \forall i\in\sN
\end{align}
where $w_i$ denotes the nodal carbon intensity at Node $i$, $R_i^{\mathrm{in}}$ and $P_i^{\mathrm{in}}$ denote the total carbon inflow and the total power inflow of Node $i$, respectively, $\sN_i^+$ denotes the set of neighboring nodes that send power to Node $i$,  $\sG_i$ denotes the set of generators at Node $i$, $p_{i,g}^{\rG}$ denotes the active power generation of the generator $g$ at Node $i$, and $w_{i,g}$ denotes the generation carbon emission factor of generator $g$ at Node $i$. Subsequently, a generic carbon-aware optimal power flow (C-OPF) model is developed based on \eqref{eq:mci} \cite{chen2023carbonawareoptimalpowerflow}. The C-OPF enables the co-optimization of both power flows and carbon flows for the optimal management of carbon emissions and power systems.  The virtual carbon-constrained network model is a representative environmental model and a powerful tool for optimizing and decarbonizing the operation of DERs. 

\paragraph{Other coupled network models}
Other geo-related network models, such as gas \cite{wang2023optimal,sawas2020resiliency}, water \cite{di2014water,zamzam2018optimal}, and transportation network models \cite{qiu2020location,unterluggauer2022electric}, are also commonly coupled  with the power system networks. Optimizing the usage of controllable grid-tied assets across different networked systems shows great promise in enhancing power grid operation, contributing to more flexible and resilient integrated power and energy systems. Note that this paper focuses on power system network models, without further detailing geo-related network models.

\subsubsection{Constraints}

\paragraph{Network constraints}

Power system network constraints ensure standard power system operations when providing reliable electricity. The management of DERs must align with grid-level constraints, such as current, voltage, and thermal constraints.  
For example, the voltage constraint can be expressed as: 
\begin{equation}
\underline{v} V_{0} \leq V_i \leq \bar{v} V_{0}, \qquad \forall i\in\sN \label{eq_network_vol}
\end{equation}
which requires that the voltage magnitudes of all nodes must be constrained within the range  $\left[\underline{v}V_{0}, \bar{v}V_{0}\right]$, $\underline{v}$ and $\overline{v}$ represent the lower and upper bounds, respectively. 

Similarly, the current constraint can be written into \cite{shchetinin2018construction}: 
\begin{equation}
\underline{\mathcal{I}}_{ij} \leq  \mathcal{I}_{ij} \leq \overline{\mathcal{I}}_{ij}, \qquad \forall ij \in \mathcal{E}\label{eq_network_current}
\end{equation}
where $\underline{\mathcal{I}}_{ij}$ and $\overline{\mathcal{I}}_{ij}$ represent the lower and upper current bounds, respectively. 

The carbon flow model also introduces networked constraints on carbon emission capacity, which can be imposed at the nodal level by: 
% \begin{equation}
% w_{i} \leq \bar{w}_{i}, \qquad \forall i \in \mathcal{N}
% \end{equation}
\begin{equation}\label{eq_network_carbon}
 w_i  P_{i} \leq \bar{R}_{i}, \qquad \forall i\in\sN
\end{equation} 
where $\bar{R}_{i}$ denotes the nodal emission capacity of Node $i$. 

% Other carbon flow constraints, such as the requirement on carbon emissions fairness and equity   \cite{sun2017analysis}, can also be employed.

\paragraph{Local constraints of DERs}

In addition to network constraints that can reflect the joint impacts of DERs, local constraints account for DER's individual operational requirements. For example, the operation of ESSs is subject to a set of local constraints, including state of charge (SoC) bounds, charging/discharging power limits, and energy efficiency constraints. For renewable supplies such as solar and wind power, they are often constrained by the maximum available energy. Besides, the carbon constraint can also be posted on the DER side to limit the emissions. 

% \hspace{0mm}\revfirst{{\bf{REV 1.2}}}{\color{blue} 
% 
DERs, such as solar PVs or ESSs, often provide energy or manage energy distribution to maintain comfort levels in buildings, particularly through HVACs.
The operation of an HVAC system can also be interpreted as a type of thermostatically controlled load (TCL) that couples thermal dynamics (e.g., insulation, ambient temperature)  and electricity consumption (e.g., \textsc{On} and \textsc{Off} cycles). 
For example, an HVAC can affect the indoor temperature according to the following linear dynamics \cite{li2011optimal}:
\begin{equation}
    \theta_{t}^{\mathrm{in}} = \theta_{t\!-\!1}^{\mathrm{in}} + \alpha_h(\theta_{t}^{\mathrm{out}}- \theta_{t\!-\!1}^{\mathrm{in}}) +\beta_h x_{t}^\mathrm{H}
\end{equation}
where $x_{t}^\mathrm{H}$ denotes the active power consumption of the HVAC at time $t$, $\theta_{t}^{\mathrm{in}}$ and $\theta_{t}^{\mathrm{out}}$ denote the indoor and outdoor temperatures at time $t$, respectively.
The parameters $\alpha_h$ and $\beta_h$ define the heat transfer properties of the environment and the thermal efficiency of the TCL appliance, respectively. 
The comfort zone constraint defines that the average temperature of the building should remain within the range of:
\begin{equation}
   \underline{\theta} \leq   {\bm{\theta}}^\mathrm{in} \leq \overline{\theta}
   \label{17s}
\end{equation}
where ${\bm{\theta}}^\mathrm{in} \in \mathbb{R}^{T}$ denotes the indoor room temperature across $T$ time intervals,  $\underline{\theta}$ and $\overline{\theta}$ denote the lower and upper bounds of the comfort zone, respectively.
% 
% }

Without loss of generality, we describe individual constraints of the $\hat{\imath}$th DER via a feasible set $\mathcal{X}_{\hat{\imath}}$, defined as: 
\begin{equation} \label{eq_individual_constraint}
\mathcal{X}_{\hat{\imath}} := \{ \hat{\imath} \in \hat{\sN}   \ |\   \underline{x}_{\hat{\imath}}  \leq x_{\hat{\imath}} \leq \bar{x}_{\hat{\imath}} \}
\end{equation}
% 
% \begin{equation} \label{eq_individual_constraint}
%    \{ x_{\hat{\imath}} \in \mathcal{X}_{\hat{\imath}} \ |  \  \underline{x}_{\hat{\imath}}  \leq x_{\hat{\imath}} \leq \bar{x}_{\hat{\imath}} \}
% \end{equation}
% 
where $x_{\hat{\imath}}$ denotes decision variable of the $\hat{\imath}$th DER and $\hat{\sN}$ denotes the set of DERs.
% The local constraints of DERs, along with any additional system-specific requirements, are essential in achieving safe, reliable, and efficient DER optimization and control.

\subsubsection{Objective functions}
The effective integration of DERs can achieve various cooperative and/or competitive grid objectives. Here, we introduce and classify these objective functions into two types, namely \emph{cooperative grid-level objectives} and \emph{competitive DER-level objectives}, highlighting the multi-faceted roles of DERs in the power system operations. 

We summarize both cooperative and competitive objectives in a general quadratic formulation as: 
\begin{equation} \label{quadra_level_obj}
f_{\mathrm{quad}}(\bm{x}) =  
    a_1\left\|\bm{A} \bm{x}+\bm{P}_t\right\|_{2}^{2} + {\bm{C}}^{T} \bm{x} + a_2
\end{equation}
where $\bm{x}_{\hat{\imath}} \in \mathbb{R}^{T}$ denotes the decision variable of the ${\hat{\imath}}$th agent expanded across $T$ time slots, $\bm{x} = [\bm{x}_1^{\mathsf{T}}, \ldots, \bm{x}_{\hat{n}}^{\mathsf{T}}]^{\mathsf{T}} \in \mathbb{R}^{\hat{n}T}$, $\hat{n}$ denotes the total number of agents, $a_1$ and $a_2$ denote cost parameters for adjusting objective weights, and $\bm{A} \in \mathbb{R}^{T \times \hat{n}T}$, $\bm{P}_t \in \mathbb{R}^{T}$, $\bm{C} \in \mathbb{R}^{\hat{n}T}$
 denote parameter matrices. The quadratic objective in \eqref{quadra_level_obj} is applicable for various power system applications, such as load shifting \cite{li2020optimal}, voltage regulation \cite{robbins2015optimal}, and EV charging control problems \cite{gan2012optimal,liu2017decentralized}.

% and $\bm{A} \in \mathbb{R}^{T \times \hat{n}T}$ denotes a constant aggregation matrix.
\paragraph{Cooperative grid-level objective functions} 
It refers to controlling DERs to improve the gird operations such as for peak shaving, valley filling, voltage regulation, frequency control, and demand response. For example, the load-shaping objective takes the form of:
\begin{align} 
f_{\mathrm{shape}}(\bm{x}) = \frac{1}{2}\left\|\bm{A} \bm{x}+\bm{P}_t\right\|_{2}^{2}
\label{eq_co1}
\end{align}
where the physical interpretation of the vector $\bm{P}_{t} \in \mathbb{R}^T$ can represent the baseline load in valley-filling problems.  

% while in the power trajectory tracking case, $\bm{P}_{t}$ is the difference between the baseline load and target profile.  

Some other cooperative grid-level objectives like frequency control and voltage regulation aim to keep the power system's frequency or voltage close to its nominal values. For example, the voltage regulation objective minimizes the squared deviation of the bus voltage magnitude by \cite{salgado2012optimal}:
\begin{equation}\label{eq_co2}
 f_{\mathrm{voltage}} (\bm{x}) = \sum_{i \in \sN} ||V_i (\bm{x}) - \hat{V}_i (\bm{x})||_2^2
\end{equation}
where $\hat{V}_i (\bm{x})$ denotes the nominal voltage magnitude output of bus $i$, which depends on the decision variable $\bm{x}$ from all agents.

% denotes the output for nominal voltage magnitude of bus $i$ that depends on the decision variable $\bm{x}$. 

%\begin{align} 
%v(\bm{x}) = \frac{1}{2}\left\|\bm{P}_{b} + \bm{B} \bm{x}\right\|_{2}^{2}
%\label{eq_1}
%\end{align}
%where $\bm{P}_{b} \in \mathbb{R}^T$ denotes the baseline load profile, $T$ denotes the total number of time slots, and $\bm{B} \in \mathbb{R}^{T \times T}$ denotes a constant matrix for the aggregation of the decision variable $\bm{x} = [x_1, \ldots,x_n]^{\mathsf{T}} \in \mathbb{R}^T$. The valley filling objective in \eqref{eq_1} has been widely adopted in PES applications, such as the EV charging problems in \cite{gan2012optimal,liu2017decentralized} that aggregate the charging loads of EVs for valley filling. 

Power loss minimization is another cooperative grid-level objective that closely relates to the grid's power flow model. The supplies and demands from DERs are flexible and can be adjusted to reduce power losses. For instance, the active power loss can be represented by \cite{mahmoud2015optimal}:
\begin{align}\label{eq_co3}
 f_{\mathrm{active}}(\bm{x}) = \sum_{l_{ij} \in \mathcal{L}} r_{ij} \left
(\frac{ \mathcal{P}_{ij}^2(\bm{x}) + \mathcal{Q}_{ij}^2(\bm{x})}{V_i^2(\bm{x})}\right).
\end{align}
where $\mathcal{P}_{ij}(\bm{x})$ and $\mathcal{Q}_{ij}(\bm{x})$ denote the active and reactive power flow outputs of the line $l_{ij}$, respectively.

% \begin{subequations} \label{4s}
% \begin{align}
%  f^{\mathrm{active}} &= \sum_{l_{ij} \in \mathcal{L}} r_{ij} \left
% (\frac{ \mathcal{P}_{ij}^2 + \mathcal{Q}_{ij}^2}{V_i^2}\right) \label{4a_ss}\\
%    f^{\mathrm{reactive}} &= \sum_{l_{ij} \in \mathcal{L}} x_{ij} \left
% (\frac{ \mathcal{P}_{ij}^2 + \mathcal{Q}_{ij}^2}{V_i^2}\right).\label{4b_ss}
% \end{align}
% \end{subequations} 

Besides, the environmental objective functions, such as minimization of CO$_2$ emissions, can be expressed as \cite{ren2010multi}:
\begin{equation} \label{eq_co5}
 f_{\mathrm{env}}(\bm{x}) = c_sp^{\mathrm{grid}}(\bm{x}) + \sum_{i \in \mathcal{N}} \sum_{u \in \hat{\mathcal{N}}_{i}} g_{i,u} p_{i,u}^{\mathrm{fuel}}(\bm{x}_{u})
\end{equation} 
where $p^{\mathrm{grid}}(\bm{x})$ denotes the total consumer power of grid electricity, multiplied by $c_s$ that denotes the carbon intensity of the grid electricity, $\hat{\mathcal{N}}_{i}$ denotes the set of agents connected to bus $i$, $p_{i,u}^{\mathrm{fuel}}(\bm{x}_{u})$ is the consumed fuels from other DER and non-DER sources,  $\bm{x}_{u}$ denotes the decision variables of the $u$th fuel source, and $g_{i,u}$ denotes the carbon intensity of the specific fuel $u$ at bus $i$.

% that can be written into the quadratic objective in \eqref{load_shaping_obj}.  

\paragraph{Competitive DER-level objective functions} DERs have objective functions based on their distinct physical properties, operational requirements, and end-user needs. These types of objective functions, i.e., $f_{\hat{\imath}}(\bm{x}_{\hat{\imath}})$, are referred to as competitive because they reflect the interest of an individual agent associated with a specific DER and involve only one decision variable (or a group of DERs acting as a single agent).

The quadratic objective in \eqref{quadra_level_obj}
also applies to a wide range of competitive DER-level objectives. For example, ESSs often suffer from battery degradation caused by frequent charging and discharging of batteries over time. The minimization of battery degradation cost is essential for improving the batteries' energy efficiency and reliability, frequently required in plug-in EVs \cite{ma2015distributed,ahmadian2017cost} and off-grid power systems \cite{bordin2017linear}. To this end, the following battery degradation cost objective can reduce the charging and discharging cycles by \cite{liu2017decentralized,ma2015distributed}: 
\begin{equation} \label{Battery_degradation_cost} 
f_{\mathrm{battery}} (\bm{x}_{\hat{\imath}}^\mathrm{b}) = \| \bm{x}_{\hat{\imath}}^\mathrm{b} \|_{2}^{2}
\end{equation}
where $\bm{x}_{\hat{\imath}}^\mathrm{b} \in \mathbb{R}^T$ denotes the charging/discharging profiles of the $\hat{\imath}$th ESS over $T$ time slots. Note that the battery degradation cost objective can also involve other factors, such as the battery's depth of discharge, the ambient temperature, and the maintenance cost \cite{abdulla2016optimal}. Similarly, capacitors and regulators are also often penalized by frequent switching control costs to slow the devices from wearing out \cite{fan2022powergym}. 

% \hspace{0mm}\revfirst{{\bf{REV 1.6}}}{\color{blue}
The comfort to users based on electrical usage of common appliances, such as washing machines and HVAC systems, is also commonly considered a DER-level objective.
The \textsc{On}-\textsc{Off} status of an HVAC  can be controlled by end users to maintain room temperature within a specific comfort zone, i.e., $[\underline{\theta},\overline{\theta}]$. 
To this end, the indoor room  temperature comfort objective can be enforced by: 
\begin{equation} \label{comfort_cost} 
f_{\mathrm{HVAC}} (\bm{x}_{\hat{\imath}}^\mathrm{H}) =  \|  \bm{\theta}^{\mathrm{in}} - \hat{\bm{\theta}}^{\mathrm{in}} \|_2^2
\end{equation}
where $\hat{\bm{\theta}}^{\mathrm{in}} \in 
\mathbb{R}^{T}$ denotes the desired room temperature across $T$ time intervals, e.g., the averaged temperature of $(\underline{\theta} + \overline{\theta})/2$. 

Washing appliances, like dishwashers and clothes washers, provide comfort to users when tasks are completed by a specific time \cite{li2011optimal}. 
In other words, the cumulative power consumption of such machines must reach a certain energy threshold by the deadline, defined by: 
\begin{equation} \label{comfort_cost_washing} 
f_{\mathrm{machine}} (\bm{x}_{\hat{\imath}}^\mathrm{M}) =  |  \sum_{t=1}^{T}(\bm{x}_{\hat{\imath},t}^\mathrm{M} \cdot \Delta t)  - \hat{e}^{\mathrm{M}}_{T} |^2
\end{equation}
where $\bm{x}_{\hat{\imath}}^\mathrm{M} \in 
\mathbb{R}^{T}$ denotes the active power consumption of the $\hat{\imath}$th washing machine across $T$ time intervals, $\Delta t$ denotes the time interval length, and $\hat{e}^{\mathrm{M}}_{T}$ denotes the desired energy level at the end of the period. 
% }

% The capacitor's switching control cost is defined as:  
% \begin{equation} \label{capacitors_regulators_cost} 
% f_{\mathrm{capacitor}} = \| \bm{x}^{c}_{t+1} - \bm{x}^{c}_{t}  \|_{2}^{2},
% \end{equation}
% where 

Another exemplary DER-level objective is the minimization of operational curtailment costs. For example, solar PV curtailment is usually framed as a loss that should be discouraged from grid and market customs \cite{o2020too}. The curtailment cost of a solar PV can be calculated based on the inverter's active and reactive power generations by  \cite{su2014optimal,wang2015dynamic}:
\begin{equation}\label{PV_curtail_cost} 
   f_{\mathrm{curtail}} (\bm{x}^{\mathrm{pv}}_{\hat{\imath}}) = ||\bm{x}^{\mathrm{pv}}_{\hat{\imath}} - \bar{\bm{x}}^{\mathrm{pv}}_{\hat{\imath}}||_2^2 + f^{\mathrm{pvg}}(\bm{s}^{\mathrm{pv}})
\end{equation}
where $\bm{x}^{\mathrm{pv}}_{\hat{\imath}} \in \mathbb{R}^T$ and $\bar{\bm{x}}^{\mathrm{pv}}_{\hat{\imath}} \in \mathbb{R}^T$ denote the curtailed and original active power generation from the solar PV, respectively, and $f^{\textsc{pvg}}$ ($\bm{s}^{\mathrm{pv}}$) denotes the solar PV generation cost that can be described via a polynomial of the apparent power $\bm{s}^{\mathrm{pv}} \in \mathbb{R}^T$, e.g., $f^{\mathrm{pvg}}(s^{\mathrm{pv}}_{t}) = c_1^{\mathrm{pv}}(s^{\mathrm{pv}}_{t})^2 + c_2^{\mathrm{pv}}s^{\mathrm{pv}}_{t} + c_3^{\mathrm{pv}}$, whose coefficients $c_1^{\mathrm{pv}}$, $c_2^{\mathrm{pv}}$, and $c_3^{\mathrm{pv}}$ can determined by curve fitting from the manufacturer. Broadly, curtailment cost objectives can be viewed as driving the states of grid-tied devices to desired values, similar to the requirement of returning a battery to its initial SoC at the end of the working period.
Additionally, the competitive DER-level objectives also include the aggregated decision-making for a group of DERs, such as the bidding plans from distribution companies and DER aggregators \cite{xiao2020coordination} and the negotiation on locational marginal price from multiple prosumers  \cite{attarha2019affinely}.

% Specifically, 

% \paragraph{\color{magenta}Fuel consumption minimization: wind, renewable energies}

% \paragraph{\color{magenta}State estimation}

% Apart from grid-level objectives, agent-level objectives

\subsection{General Problem Formulation}

%After identifying the constraints and objective functions, it is ready to frame them into a constrained PES optimization problem. The formulated optimization problems can be categorized by the coupling of decision variables in the objective functions and constraints. 

% In what follows, we present two exemplary problem formulations-- one consists of separable objectives and constraints and one in the general coupled optimization problem formulation.

After identifying the objective functions and constraints, we present the mathematical formulation of the DER control problem as: 
\begin{pequation} \label{general_optimization_problem}
\begin{aligned}
& \min_{\bm{x}} && {\sum_{{\hat{\imath}} \in \mathcal{I}} f_{\hat{\imath}}(\bm{x}_{\hat{\imath}})} + g(\bm{x}) \\
& \: \, \suchthat  && \bm{x}_{\hat{\imath}} \in \mathcal{X}_{\hat{\imath}}, \ \forall {\hat{\imath}} \in \mathcal{I}\\
 &  &&\bm{x} \in \mathcal{G}. 
\end{aligned}
\end{pequation}Problem \eqref{general_optimization_problem} aligns with \eqref{DER_problem} where the ${\hat{\imath}}$th agent is associated a decision variable $\bm{x}_{\hat{\imath}} \in \mathbb{R}^{T}$ and an objective function $f_{\hat{\imath}}(\cdot): \mathbb{R}^T \mapsto \mathbb{R}^1$, $\mathcal{I}$ denotes the set of agents, $T$ denotes the dimension, $\mathcal{X}_{\hat{\imath}}$ denotes a compact set that describes the feasible region of the decision variable $\bm{x}_{\hat{\imath}}$, $\bm{x} = [\bm{x}_1^{\mathsf{T}}, \ldots, \bm{x}_{\hat{n}}^{\mathsf{T}}]^{\mathsf{T}}$, $g(\cdot):\mathbb{R}^{\hat{n}T} \mapsto \mathbb{R}^1$ denotes a coupled objective function whose inputs are collected decision variables from all agents, and $\mathcal{G}$ denotes a convex set that describes the  coupled constraints including the network model and network constraints.

Problem \eqref{general_optimization_problem} has been broadly adopted to optimize the operation of power electric systems, such as demand response \cite{diekerhof2017hierarchical}, optimal power flow \cite{peng2014distributed}, management of grid-interactive efficient buildings \cite{yu2023online}, and EV charging control problems \cite{rivera2016distributed}. Problem \eqref{general_optimization_problem} represents a generalized DER control problem formulation, containing coupled objective functions and constraints (e.g., \eqref{eq_co1}-\eqref{eq_co5} and \eqref{eq_network_vol}, \eqref{eq_network_carbon}), 
and separable objective functions and constraints (e.g., \eqref{Battery_degradation_cost},\eqref{PV_curtail_cost}, and \eqref{eq_individual_constraint}).  
Additionally, it has also been broadly applied in other industrial cyber-physical system applications, such as rate control in communication networks \cite{kelly1998rate}, coordination of connected and autonomous vehicles \cite{tajalli2018distributed}, path tracking of unmanned aerial vehicles \cite{mansouri2015distributed}, control of nonlinear systems \cite{pierer2023asynchronous}, and congestion management in transportation systems \cite{he2007towards}.

\section{Scalable Methods}
\label{Scalable_Methods}

This section reviews state-of-the-art and fundamental scalable algorithms within multi-agent frameworks. We select representative works under each category of scalable multi-agent approaches and show their applications in DER control with highlighted key features (see Table \ref{table_scalable_alg}). At the end, we discuss related privacy leakage issues in these typical multi-agent frameworks for DER control. 

%The DER control problem, e.g., Problem \eqref{general_optimization_problem} based on \eqref{DER_problem}, can be solved via distributed or decentralized information exchange structures.  Therefore, we first present a variety of scalable methodologies that are built upon distributed and decentralized architectures, then survey their applications in DER control problems. 

\subsection{Distributed and Decentralized Algorithms}

\subsubsection{Average consensus} 
Average consensus (AvgC) includes \emph{dynamic AvgC}, where agents seek to compute the average of individual time-varying signals, and \emph{static AvgC}, where agents reach the average of their initial values. The convergence of AvgC is first proved by DeGroot \cite{degroot1974reaching}, then further studied by many researchers (see, e.g., \cite{colfati2007consensus,olshevsky2009convergence,horn2012matrix}). 
To provide a straightforward explanation, we refer to AvgC as the static one and introduce its theoretical foundations. AvgC-based algorithms are commonly used in multi-agent systems to collaboratively compute the average of agents' local values.  Suppose each agent has an initial scalar state $x_i^0$. The average consensus asymptotically converges to an ``agreement,'' e.g., a constant $c$, under suitable assumptions on the coefficients and graph connectivity. At the $\ell$th iteration, agent $\hat{\imath}$ updates its decision variable $x_{\hat{\imath}}^{\ell} \rightarrow x_{\hat{\imath}}^{\ell+1}$ by \cite{degroot1974reaching, olshevsky2009convergence}:
\begin{equation} \label{consensus_original}
x_{\hat{\imath}}^{\ell+1}=\sum_{\hat{\jmath}=1}^{\hat{n}} a_{\hat{\imath} \hat{\jmath}}^{\ell}x_{\hat{\jmath}}^{\ell}
\end{equation}
where $x_{\hat{\imath}}^{\ell+1}$ is the weighted average held by the agent $\hat{\imath}$, $a_{\hat{\imath} \hat{\jmath}}^{\ell}$
denotes the averaging coefficient. Follow \eqref{consensus_original}, the averaged consensus is achieved at $\lim _{\ell \rightarrow \infty} x_{\hat{\imath}}^{\ell} = c, \forall \hat{\imath} \in \mathcal{I}$. 

Based on the distributed multi-agent information exchange structure, the $\hat{\imath}$th agent can achieve AvgC by interacting only with its neighbors as \cite{colfati2007consensus}:
\begin{equation} \label{consensus}
x_{\hat{\imath}}^{\ell+1}=x_{\hat{\imath}}^{\ell}+\epsilon \sum_{\hat{\jmath} \in \mathcal{B}_{\hat{\imath}}} \phi_{\hat{\imath} \hat{\jmath}}\left(x_{\hat{\jmath}}^{\ell}- x_{\hat{\imath}}^{\ell}\right)
\end{equation} where $\mathcal{B}_{\hat{\imath}}$ denotes the set of neighbors of agent $\hat{\imath}$, $\epsilon$ denotes the step size, and $\phi_{\hat{\imath} \hat{\jmath}}$ denotes the adjacency coefficient of the network, i.e., $\phi_{\hat{\imath} \hat{\jmath}} = 0$ if $\hat{\jmath} \notin \mathcal{B}_{\hat{\imath}}$. Follow the distributed information exchange structure, the decision variables $x_{\hat{\imath}}^{\ell}$, $\forall \hat{\imath} \in \mathcal{B}$ converge to the averaged value $c=\sum_{\hat{\imath}=1}^{\hat{n}} x_{\hat{\imath}}^0/\hat{n}$, under balanced digraph and other numerical assumptions (see more details in \cite{horn2012matrix, colfati2007consensus}). 
% \revfirst{{\bf{REV 1.7}}}{\color{blue} 
The scalable architecture of distributed AvgC enables cooperative multi-agent decision-making \cite{ren2005survey,mehyar2005distributed,soatti2016consensus,cybenko1989dynamic,yang2013consensus,li2017consensus,binetti2014distributed,pourbabak2017novel,oh2015survey}. These AvgC methods are widely utilized in diverse fields that pertinent to multi-agent-based DER control problems, including networked control theory \cite{jadbabaie2003coordination}, communication networks \cite{mehyar2005distributed,soatti2016consensus}.
For smart grid applications, AvgC-based methods have gained prominence, especially in the control of grid-tied energy resources \cite{cybenko1989dynamic,yang2013consensus,li2017consensus,binetti2014distributed,pourbabak2017novel,rana2017consensus}.
The unique trait of AvgC that ensures that all agents can reach an agreement based on their initial values/opinions offers the opportunity 
for well-suited deployment of DER control problems, such as achieving optimal DER management for supply-demand balance \cite{xu2014distributed,khazaei2016consensus,huang2017distributed,li2017consensus}.
% 
% }
% \revfirst{{\bf{REV 1.8}}}{\color{blue}
Therefore, AvgC is efficient for distributed coordination and primarily used for tasks like distributed averaging, decentralized estimation, and synchronization in networked multi-agent systems.
However, AvgC-based methods are often limited to scenarios where all agents must reach the same consensus value.
Additionally, future research is needed to accelerate convergence in large networks or those with weak connectivity and to reduce sensitivity to time delays and changes in network topology.
% 
% }
% 
Variations of AvgC-based algorithms have been developed to handle asynchronous and time-varying environments \cite{tsitsiklis1984problems,tsitsiklis1986distributed,ren2005consensus}, accelerate convergence through linear predictors \cite{aysal2008accelerated}, and incorporate quantization techniques that refine intervals as the algorithm progresses \cite{thanou2012distributed}.
%
% \revfirst{{\bf{REV 1.7}}}{\color{blue} 
These new techniques are crucial for the research and development in the power and energy field, as power systems inherently exist as networked systems in dynamic environments.
% }

\subsubsection{Alternating direction method of multipliers} 

Alternating direction method of multipliers (ADMM) is initially developed in \cite{gabay1976dual} based on the augmented Lagrangian and later independently rediscovered and popularized by Boyd \emph{et al.} \cite{boyd2011distributed}. ADMM has been popular in optimizing large-scale multi-agent systems  owing to its decomposition ability. Specifically, it focuses on solving a type of optimization problem:

\begin{pequation} \label{ADMMp3}
\begin{aligned}
&\underset{\Tilde{\bm{x}},\Tilde{\bm{y}}}{\text{min}} & & f(\Tilde{\bm{x}}) + g(\Tilde{\bm{y}})  \\
& \,  \suchthat & & \bm{D}\Tilde{\bm{x}} + \bm{G}\Tilde{\bm{y}} = \bm{h} 
\end{aligned}
\end{pequation}where $\Tilde{\bm{x}} \in \mathbb{R}^{T_1}$ and $\Tilde{\bm{y}} \in \mathbb{R}^{T_2}$ are variables, $\bm{D} \in \mathbb{R}^{m \times T_1}$ and $\bm{G} \in \mathbb{R}^{m \times T_2}$ are two matrices, and $\bm{h} \in \mathbb{R}^{m}$ is a $m$-dimensional vector. The objective functions, $f(\cdot)$ and $g(\cdot)$, are assumed to be convex.

ADMM forms an augmented Lagrangian of \eqref{ADMMp3} as: 
\begin{align}
    \mathcal{L}_\rho(\Tilde{\bm{x}}, \Tilde{\bm{y}}; \bm{\lambda}) &= f(\Tilde{\bm{x}}) + g(\Tilde{\bm{y}}) + \bm{\lambda}^{\mathsf{T}} ( \bm{D}\Tilde{\bm{x}} + \bm{G}\Tilde{\bm{y}} - \bm{h})
    \nonumber\\
     & \qquad + \frac{\rho}{2} \| \bm{D}\Tilde{\bm{x}} + \bm{G}\Tilde{\bm{y}} - \bm{h}\|_2^2
\label{Lagrangian}
\end{align}
where $\bm{\lambda} \in \mathbb{R}^{m}$ denotes the Lagrange multiplier associated with the equality constraint, and $\rho>0$ denotes the penalty parameter associated with the penalty term $\| \bm{D}\Tilde{\bm{x}} + \bm{G}\Tilde{\bm{y}} - \bm{h}\|_2^2$. The penalty term, or regularization, adds an extra cost to the optimization function, penalizing the model when it deviates from the constraint.

Based on the augmented Lagrangian, the ADMM updates the primal (decision variable) and the dual variable (Lagrange multiplier) by:
\begin{subequations}\label{ADMM updates}
\begin{align} \Tilde{\bm{x}}^{\ell+1} & =\underset{\Tilde{\bm{x}}}{\operatorname{argmin}} ~  \mathcal{L}_{\rho}\left(\Tilde{\bm{x}}, \Tilde{\bm{y}}^{\ell}; \bm{\lambda}^{\ell}\right) \\ 
\Tilde{\bm{y}}^{\ell+1} & =\underset{\Tilde{\bm{y}}} {\operatorname{argmin}}~\mathcal{L}_{\rho}\left(\Tilde{\bm{x}}^{\ell+1}, \Tilde{\bm{y}}; \bm{\lambda}^{\ell}\right) \\ 
\bm{\lambda}^{\ell+1} & = \bm{\lambda}^{\ell}+\rho\left(\bm{D}\Tilde{\bm{x}}^{\ell+1} + \bm{G}\Tilde{\bm{y}}^{\ell+1} - \bm{h}\right).\end{align}
\end{subequations} 

Since $f(\Tilde{\bm{x}})$ and $g(\Tilde{\bm{y}})$ have uncorrelated decision variables, the decomposability of ADMM allows  $\Tilde{\bm{x}}$ and $\Tilde{\bm{y}}$  to be updated separately  in a sequential (alternating) fashion. 
% \revfirst{{\bf{REV 1.8}}}{\color{blue}
% 
The distributed nature of ADMM enables scalability in solving large-scale multi-agent optimization problems. By decomposing large problems into smaller sub-problems, ADMM-based approaches allow individual agents to solve these sub-problems locally. They are suitable for distributed optimization problems such as distributed machine learning \cite{gebbran2022multiperiod}, large-scale networked systems \cite{ling2016weighted}, and resource allocation problems in power grids and communication networks \cite{attarha2019affinely}.
Despite the merits, ADMM-based methods can suffer from high communication overhead between agents due to the need for iterative message exchanges. 
% Other promising directions include improving convergence speed in poorly conditioned problems and 
% implementations in asynchronous systems.
% 
% }
Recently, ADMM has been extensively studied and improved with a number of generalizations, including approaches for tackling nonseparable optimization problem formulations \cite{chen2019extended,mihic2021managing}, nonconvex problems \cite{wang2019global,themelis2020douglas,lu2021linearized}, as well as other heuristic ADMM-based approaches \cite{takapoui2020simple,diamond2018general}. Therefore, ADMM-based methods are well-suited  for solving large-scale DER optimization problems, including the management of DERs with high uncertainty of power generation and load forecasts \cite{ma2016distributed,attarha2019affinely},
the decomposition of OPF with non-linear and non-convex formulations
\cite{erseghe2014distributed,magnusson2015distributed}, and asynchronous distributed optimization algorithms \cite{zhang2016admm}.

\subsubsection{Projected gradient descent} \label{PGD} 
Gradient descent is a fundamental method to solve unconstrained optimization problems. Gradient descent iteratively moves towards the minimum of a function by taking steps proportional to the negative of the gradient of the function. Compared to gradient descent, projected gradient descent (PGD) uses additional projection operations to enforce constraints by projecting the solution back into the feasible region after updating primal and/or dual variables.  PGD-based methods are well suited in solving constrained optimization problems, particularly for large-scale optimization tasks with numerous local constraints and continuously differentiable objective functions.

For example, the relaxed Lagrangian function of  problem \eqref{general_optimization_problem} is:
\begin{align}
    \mathcal{L}_r(\bm{x}; \bm{\lambda}) = {\sum_{{\hat{\imath}} \in \mathcal{I}} f_{\hat{\imath}}(\bm{x}_{\hat{\imath}})} + g(\bm{x}) + \bm{\lambda}^{\mathsf{T}} ( \bm{A}\bm{x} - \bm{h})
\label{Lagrangian_PGD}
\end{align}
where $\sG$ in \eqref{general_optimization_problem} is defined as $\sG \coloneq \bm{A}\bm{x} - \bm{h} \leq \bm{0}$.

Subsequently, PGD updates the primal \eqref{11a_ss} and dual \eqref{11b_ss} variables by
\cite{bertsekas2015parallel}:
\begin{subequations} \label{eq1}
\begin{align}
    \bm{x}_{\hat{\imath}}^{\ell+1} &= \Pi_{\mathbb{X}_{\hat{\imath}}}[\bm{x}_{\hat{\imath}}^{\ell} - \alpha_{\hat{\imath}} \nabla_{\bm{x}_{\hat{\imath}}} \mathcal{L}_r\left(\bm{x}_{1}^{\ell},\ldots,\bm{x}_{\hat{n}}^{\ell}; \bm{\lambda}^{\ell}\right)] \label{11a_ss}\\
     \bm{\lambda}^{\ell+1} &=\Pi_{\mathbb{R}^{+}}[\bm{\lambda}^{\ell} + \beta\nabla_{\bm{\lambda}} \mathcal{L}_r \left(\bm{x}_{1}^{\ell},\ldots,\bm{x}_{\hat{n}}^{\ell}; \bm{\lambda}^{\ell}\right)] \label{11b_ss}
\end{align}
\end{subequations} where $\alpha_{\hat{\imath}}$ and $\beta$ denote the primal and dual step sizes, respectively, $\mathcal{L}_{r}(\bm{x}_{1}^{\ell},\ldots$, $\bm{x}_{\hat{n}}^{\ell}; \bm{\lambda}^{\ell})$ denotes the relaxed Lagrangian function at the $\ell$th iteration, $\Pi_{\mathcal{X}_{\hat{\imath}}}[\cdot]$ denotes the Euclidean projection operator, and $\mathbb{R}^{+}$ denotes the positive real set. Regularization terms, such as the penalty $\frac{\rho}{2} \| \bm{D}\Tilde{\bm{x}} + \bm{G}\Tilde{\bm{y}} - \bm{h}\|_2^2$ in \eqref{Lagrangian}, can also be included in \eqref{Lagrangian_PGD} to enforce the satisfaction of constraints and enhance convergence  \cite{koshal2011multiuser,liu2017decentralized}. 

% \hspace{0mm}\revfirst{{\bf{REV 1.8}}}{\color{blue}
% 
% 
PGD-based methods are straightforward to implement and computationally efficient for large-scale multi-agent problems. The PGD allows agents to handle local constraints individually, making it a good fit for decentralized settings. 
However, PGD can suffer from slow convergence when addressing non-convex problems or poorly conditioned constraints, as it is sensitive to step size choices and requires careful tuning for optimal performance. Additionally, agents must perform a projection step, which can be computationally expensive for certain constraint sets.
% 
% }
To enhance scalability, generality, and convergency in optimizing multi-agent systems, PGD-based algorithms have been continuously improved to include a series of recent findings, such as the regularized primal-dual subgradient method that can deal with non-separable objectives and constraints \cite{koshal2011multiuser}, shrunken primal-dual subgradient that eliminates the regularization errors \cite{liu2017decentralized}, and shrunken primal-multi-dual subgradient that achieves two-facet scalable \emph{w.r.t.} both the network dimension and the agent population size \cite{huo2022two}. PGD-based (and gradient-based) approaches have been widely adopted for managing DER-rich  power grids, examples include solving online load flow optimization problems  \cite{hauswirth2016projected}, decentralized management of renewable generation and demand response \cite{bahrami2018decentralized}, and voltage regulation with DERs \cite{zhou2019accelerated}.

% terms, e.g., $\frac{\rho}{2} \| \bm{A}\bm{x}_1 + \bm{B}\bm{x}_2 - \bm{c}\|_2^2$
% in ADMM,

\subsubsection{Multi-agent reinforcement  learning} Learning-aided approaches, especially multi-agent reinforcement learning (MARL), are efficient for data-driven decision-making for power systems with proliferating DERs \cite{chen2022reinforcement}. Mathematically, the decision-making is formulated into a \emph{Markov Decision Process} (MDP), defined by the state space $\mathcal{S}$, action space $\mathcal{A}$, the transition probability function $\mathbb{P}(\cdot | s, a)$ that maps a state-action pair $(s, a) \in \mathcal{S} \times \mathcal{A}$ to a distribution on the state space, and the reward function $r(s,a)$. The agents aim to find an optimal policy $\pi^*$ that maximizes the expected infinite horizon discounted reward $J(\pi)$, defined by  \cite{sutton2018reinforcement}:
\begin{equation}
    \pi^{*} \in \arg \max _{\pi} J(\pi)=\mathbb{E}_{s_{0} \sim \mu_{0}} \mathbb{E}_{\pi} \sum_{t=0}^{\infty} \gamma^{t} r\left(s_{t}, a_{t}\right)
\end{equation}
where $\mathbb{E}$ denotes the expectation, $s_0$ is drawn from an initial state distribution $\mu_0$, $a_{t}$ is taken according to the policy $\pi$, 
$\gamma^{t} \in (0,1)$ denotes the discounting factor for the future rewards at time $t$. By interacting with the environment, RL agents learn the optimal policy without the knowledge of the model, i.e., via the transition probability and the reward function. 

MARL involves the interactive decision-making of multiple agents that operate in a common environment \cite{zhang2021multi,al2021adaptive}. Many power system management problems, including DER control, can be cast into the realm of MARL, where various grid components, such as generators, controllers, or local operators, can act as independent agents and operate under the grid environment. In MARL, the $\hat{\imath}$th agent takes an action $a_{t}^{\hat{\imath}} \in \sA_{\hat{\imath}}$, given the state $s_t$, and receives a reward $r_t^{\hat{\imath}}(s_{t}, \{a_{t}^{\hat{\imath}}\}_{\hat{\imath} \in \mathcal{I}})$, then the system state $s_{t}$ transits into $s_{t+1}$. 
In power system applications, the states can include currents or voltages at different buses, real and reactive power demands, line flows, the status of DERs (e.g., battery energy level), transformer tap positions, etc. The actions can be taken on adjusting active/reactive power outputs, changing transformer tap settings, switching capacitors or reactors (on/off), adjusting load shedding levels, reconfiguring network topology, etc.

% \hspace{0mm}\revfirst{{\bf{REV 1.8}}}{\color{blue}
% 
MARL-based approaches are compatible with dynamic, uncertain, and complex environments where agents learn to make decisions through interactions with the environment.
When predefined strategies fail in uncertain environments, MARL enables agents to adapt by learning from their actions and their consequences over time, capable of solving a wide range of dynamic tasks, including coordination, cooperation, and competition between agents.
% 
% }
% 
% 
Based on the information exchange pattern between agents and the SO or coordinator, 
% \revsecond{{\bf{REV 2.6}}}{\color{red}
MARL algorithms can also be categorized into three representative types as presented in Fig. \ref{central_distri_decentr}, including \textit{centralized} (also referred to as centralized training with decentralized execution), \emph{decentralized}, and \emph{distributed} structures (also referred to as decentralized setting with networked agents), see \cite{zhang2021multi} for more details. 
% }
% \deleted{Specifically: 1) \textit{Centralized}, also referred to as \emph{centralized training with decentralized execution}. It assumes the existence of a central controller that can collect data from all agents, including their actions, rewards, and observations. Having a coherent view of the entire system, centralized settings significantly simplify the analysis \mbox{\cite{foerster2017stabilising,gupta2017cooperative,lowe2017multi}}. However, centralized settings unavoidably suffer from scalability issues, such as the exponential growth of the joint action space. The scalability challenge motivates the development of decentralized or distributed structures that do not rely on a central controller; 2) \emph{Fully decentralized}, where there is no direct exchange of information between any agents. Instead, each agent makes decisions independently based on its local observations, without any coordination and/or data aggregation. Fully decentralized structures largely enhance the scalability and eliminate peer-to-peer communications. However, they could suffer from non-convergence issues or delayed learning caused by the lack of a global view; 3) \emph{Distributed} structure allows agents to communicate with others (e.g., neighbors) through a potentially time-varying communication network. Owing to the additional share of local information, distributed MARL structures benefit from better theoretical conciseness. }
Scalable MARL methods are powerful and promising  tools for controlling DERs in  large-scale power system networks with high-dimensional data streams \cite{chen2021powernet,charbonnier2022scalable,cui2022survey,zhang2020multi}. 
% \revfirst{{\bf{REV 1.8}}}{\color{blue}
However, MARL-based methods commonly face 
high computational complexity and slow convergence in large-scale systems due to the curse of dimensionality and exploration-exploitation trade-offs. Besides, coordination between agents is challenging in dynamic power systems, especially in partially observable environments.
Another future direction involves addressing the requirement for extensive training data, as learned policies may not generalize well to new scenarios, such as cyber and physical attack situations in power systems \cite{NAERC_cyber}.
% }

% Other scalable learning-based approaches, such as federated learning (FL) and transfer learning,  can also enable scalability and privacy by distributing computation across clients in conjunction with privacy preservation measures 
% \cite{collins2021exploiting}. 

%In Table \ref{table_scalable_alg}, we name only a few literature to illustrate the design of scalable multi-agent approaches and their applications in DER control. This paper focuses on reviewing the orchestrated scalability and privacy preservation methods that are presented in Tables \ref{table_DP_review}-\ref{Table_others_literature}. 

% $^\star$\footnotesize{Decryption} $^\ast$\footnotesize{Share Generation} $^\circ$\footnotesize{Secret Reconstruction}

% Transfer learning 

% Machine learning 

% GCNN

% @inproceedings{abadi2016deep,
%   title={Deep learning with differential privacy},
%   author={Abadi, Martin and Chu, Andy and Goodfellow, Ian and McMahan, H Brendan and Mironov, Ilya and Talwar, Kunal and Zhang, Li},
%   booktitle={Proceedings of the 2016 ACM SIGSAC conference on computer and communications security},
%   pages={308--318},
%   year={2016}
% }

\begin{remark} 
% \normalfont
Note that we follow the information exchange categorization in Fig. \ref{central_distri_decentr} to classify the distributed and decentralized information exchange structures. One major difference between them is that decentralized structures have no peer-to-peer communications. However, distributed and decentralized algorithms can be case-dependent when categorized into distributed and decentralized information exchange structures. For example, a distributed algorithm like ADMM can also be executed in a decentralized structure, where the coordinator collects decision variables from all the agents, and vice versa for decentralized algorithms. \hfill $\square$ 

% \textcolor{red}{my question regarding this remark is what do you want to explain or emphasize here? Does it relate to the Table 1 or the MARL algorithms? It seems like a general statement/common sense.}
%  % \qed
\end{remark}

%\begin{adjustbox}{scale=0.9}
\scriptsize
\begin{longtable}{m{3em} m{2em} m{6em} m{9.5em} m{18em}} 

\caption{Representative Work on Scalable Multi-Agent Frameworks and Their Applications in DER Control.} \label{table_scalable_alg} \\

    \toprule
    Method & Ref.  & Structure  & Applications & Key Features \\
    \midrule
    \endfirsthead
    \caption[]{(Continued)}\\
    \toprule
    Method & Ref.   & Structure  & Applications & Key Features \\
    \midrule
    \endhead
    \bottomrule
    \endfoot

    AvgC  & \cite{olshevsky2009convergence}  &  Distributed & Networked multi-agent systems & 1-Consider both fixed and time-varying topologies; 2-study convergence rate of a variety of consensus and averaging algorithms.\\

    AvgC  & \cite{huang2009coordination}  &  Distributed & Networked multi-agent systems & 1-Coordination and consensus of networked agents under noisy measurements of neighbors' states; 2-propose stochastic approximation-type algorithms with a decreasing step size; 3-introduce the notions of mean square and strong consensus.\\

    AvgC  & \cite{amelina2015approximate}  &  Distributed & Load balancing & 1-Approximate consensus problem for stochastic networks with nonlinear agents; 2-consider switching topology, noisy, and delayed information about agent states.\\
        
    AvgC  & \cite{de2018power}  &  Distributed & DC microgrids  & 1-Nonlinear consensus-like system of differential-algebraic equations; 2-controllers to converge to weighted power measurement at the sources.\\   

    \midrule

    ADMM  & \cite{ma2016distributed}  &  Distributed & Microgrids with DERs & 1-Online energy management based on ADMM; 2-explore the use of regret minimization; 3-utility microgrid buys/sells power from/to other microgrids.\\

    ADMM  & \cite{attarha2019affinely}  &  Distributed & Coordination of prosumer-owned DERs & 1-An affinely adjustable robust extension of ADMM that is resilient to forecast deviations; 2-enable prosumers to take local ``wait-and-see'' recourse decisions that compensate real-time forecast deviations.\\

    ADMM  & \cite{gebbran2022multiperiod}  & Distributed & AC optimal power flow & 1-Distributed three-block algorithm; 2-introduce carefully tuned delays in the Volt-Var control block update to circumvent unstable numerical behavior.\\

    ADMM  & \cite{mak2023learning}  &  Decentralized & AC optimal power flow  & 1-Use machine learning to speed up the convergence of ADMM; 2-develop novel data-filtering techniques to identify high-quality training data.\\

    \midrule

    PGD  & \cite{koshal2011multiuser}  &  Distributed & Multi-agent problems  & 1-Adopt Tikhonov regularization to deal with coupling objectives and constraints; 2-allow for differing step lengths across users as well as across the primal and dual space.\\

    PGD  & \cite{liu2017decentralized}  &  Decentralized & EV charging control & 1-Decentralized EV charging control for valley-filling; 2-nonseparable objective function and coupled inequality constraints; 3-develop a shrunken-primal-dual subgradient algorithm.\\

    PGD  & \cite{huo2022two_facet}  & Decentralized & Networked multi-agent systems & 1-Two-facet scalability \emph{w.r.t.} both the agent population size and the network dimension; 2-computing load reduction compared to full-dimension cases.\\

    PGD  & \cite{grimmer2024provably}  & $^\star$  & Smooth convex optimization  & 1-Prove convergence of gradient descent using nonconstant, long stepsize patterns, for smooth convex optimization; 2-computer-assisted analysis.\\

    \midrule

    MARL  & \cite{wang2023towards}  & Hierarchical (partially observable) & Mobile power sources and repair crews  & 1-Formulate a resilience-driven dispatch problem; 2-a hierarchical MARL with embedded function encapsulating system dynamics.\\

    MARL  & \cite{wang2024scalable}  & Centralized training with decentralized execution & Residential hybrid energy system & 1-A multi-stage proximal policy optimization on-policy framework with imitation learning; 2-improve indoor thermal comfort and energy efficiency.\\

    MARL  & \cite{ying2024scalable}  & Distributed training without global observability & Multi-agent problems  & 1-Safe MARL formulation that extends beyond cumulative forms in both the objective and constraints; 2-a scalable primal-dual actor-critic method.\\

    MARL  & \cite{zhang2023global}  & Distributed & Networked multi-agent problems & 1-Cooperatively maximize the average of their entropy-regularized long-term rewards; 2-localized policy iteration algorithm that provably learns a near-globally-optimal policy using only local information. \footnotemark[1]\\
    
    \bottomrule
%     \vspace{2mm}
% \footnotesize{$~~~~~~~~~~$ $^\star$Not defined in the literature ~~ AvgC: Average consensus ~~ ADMM: Alternating direction method of multipliers ~~  PGD: Projected gra- \\ ~~~~~~~~~~ dient descent ~~  MARL: Multi-agent reinforcement learning}
%\end{adjustbox}
\end{longtable}
\vspace{-0.4cm}
\scriptsize{\noindent$^\star$Not defined in the literature ~~ Ref.: Reference ~~  AvgC: Average consensus ~~ ADMM: Alternating direction method of multipliers ~~  PGD: Projected gradient descent ~~  MARL: Multi-agent reinforcement learning}
\normalsize
% \footnotetext[1]{This is the first footnote for the AvgC method.}
% Footnotes
%\end{adjustbox}

\subsection{Privacy Leakages}
\label{Privacy_Issues}

The acquisition, processing, and transmission of private customer data (e.g., energy consumption patterns, demographic data, locations, and regional statistics) are generally required to achieve grid services and improve customer satisfaction (e.g., billing, load monitoring, and demand response \cite{Gough9411689,dvorkin2020differentially,dong2018privacy,atmaca2024privacy}). However, unauthorized usage of private data can lead to privacy leakages and malicious manipulation of the system \cite{prudenzi2002neuron,tan2013increasing,hong2017privacy}, introducing vulnerabilities and thus restraining the deployment of advanced DER management approaches. 

To ensure the warranted use of private information from all stakeholders, it is crucial to synthesize  privacy preservation techniques into the design of scalable DER control strategies. Toward this goal, we summarize the typical adversaries/threats in multi-agent computing frameworks,  including external eavesdroppers, honest-but-curious agents, and the SO and/or coordinators/aggregators, each representing a distinct type of threat with different attack vectors. By examining these three adversaries, we cover a broad spectrum of external, internal, and hierarchical privacy risks, offering a comprehensive understanding of the threats in multi-agent systems. This helps guide the design of privacy-preserving multi-agent frameworks that require privacy guarantees in different operational scenarios.

%, then review existing strategies against these typical types of adversaries.

\subsubsection{External eavesdroppers}
External eavesdroppers are external adversaries who wiretap and intercept the communication channels of the power systems, e.g., data transmitted between smart meters and energy retailers. Through the acquisition of private customer and/or system information, external eavesdroppers can ``observe'' the system status and exploit system vulnerabilities without tempering the system, causing adverse effects such as financial losses, reputational damage, and operational disruptions.

\subsubsection{Honest-but-curious agents} Honest-but-curious agents, also referred to as semi-honest agents, are internal adversaries who follow the problem-solving procedures but are curious and try to infer the privacy of other participants. Being ``honest'' is the primary characteristic of this type of adversary, indicating that it must follow the prescribed procedures and cannot send any falsified message.  Despite their honest intentions, their curiosity may motivate them to steal others' private information based on their legitimately received messages and internal knowledge about the system. In contrast to external eavesdroppers, honest-but-curious agents lack the capability to intercept communication channels. However, given their role as internal participants, they present a more significant challenge to designing privacy-preserving multi-agent computing approaches. Their privileged internal access allows them to infer the private data of other participants clandestinely.

\subsubsection{The system operator and/or coordinators/aggregators} The system operator (SO)/coordinators/aggregators are usually responsible for ensuring the reliable operation of power grids. Therefore, these roles often have access to critical system information, such as network topology, protection settings, and historical demand data. Even though the SO and/or coordinators/aggregators are typically perceived as trustworthy, a dishonest or corrupted SO/coordinator/aggregator can ultimately result in the privacy compromise of the entire system. On the one hand, the SO/coordinators/aggregators may attempt to learn the DERs' decision variables by conveniently collecting and analyzing the acquired and belonged data. On the other hand, consumers and prosumers are often reluctant to disclose personal private information to any third party.

\begin{figure*}[!htb] 
\centering
\begin{subfigure}[t]{0.3\textwidth}
\centering
\includegraphics[width=1\textwidth,trim = 0mm 0mm 0mm 0mm, clip]{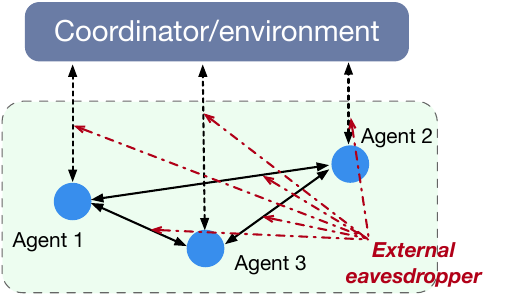}
\caption{An external eavesdropper.}
\end{subfigure}%
~ 
\begin{subfigure}[t]{0.3\textwidth}
\centering
\includegraphics[width=0.84\textwidth,trim = 0mm 3mm 0mm 0mm, clip]{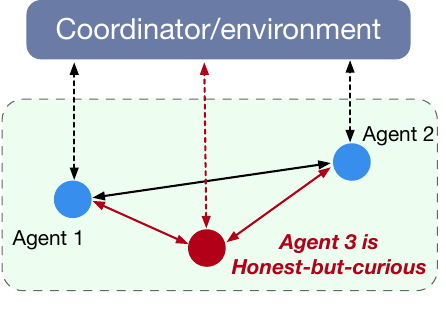}
\caption{An honest-but-curious agent.}
\end{subfigure}
~ 
\begin{subfigure}[t]{0.3\textwidth}
\centering
\includegraphics[width=0.85\textwidth,trim = 0mm 0mm 0mm 0mm, clip]{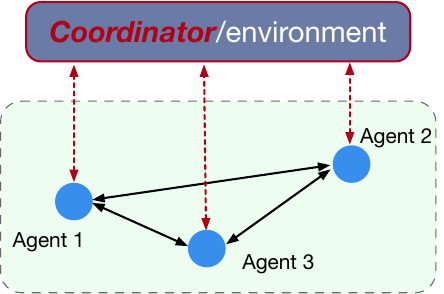}
\caption{The coordinator.}
\end{subfigure}
\caption{Illustration of privacy breaches from \emph{external eavesdroppers, honest-but-curious agents}, and \emph{the coordinator} in a three-agent distributed information exchange structure: The \emph{External eavesdroppers} wiretap all communication channels in the network; Agent three is an \emph{Honest-but-curious agent} who attempts to infer other agents' private information based on its accessible information; \emph{The Coordinator} might have access to agents' private data and/or critical system information. }
\label{fig_distrbuted_privacy_breaches}
\end{figure*}
Fig. \ref{fig_distrbuted_privacy_breaches} shows the potential privacy breaches caused by the aforementioned three types of adversaries, exemplified in a three-agent distributed information exchange structure. To summarize, privacy protection emphasizes adversarial scenarios where all participants adhere to the algorithm/protocol steps but try to get insights into the system or agent information. These adversaries are often referred to as \emph{passive} adversaries, meaning that each participant must not alter input variables or parameters and must accurately compute the outputs based on the algorithm design because they are interested in learning the correct results. Therefore, this paper focuses on reviewing scalable and privacy-preserving multi-agent frameworks in the presence of only passive adversaries. More details on \emph{passive} and \emph{active} adversaries can be referred to \emph{Remark \ref{remark_active_passive_adveraries}}.

\begin{remark} \label{remark_active_passive_adveraries}
% \normalfont
In privacy-aware and cybersecure computing, two primary types of adversaries can be categorized based on their divergence from the protocol, i.e., \emph{passive} and \emph{active} adversaries. The passive adversaries adhere to the protocol to obtain the correct results of its execution, but they also attempt to gather additional information about other participants' private information beyond what they are authorized to know. In contrast, active adversaries deviate from the protocol and tend to disrupt the computation process by modifying inputs, injecting malicious content, and tampering with intermediate results to compromise privacy or security. Compared to active adversaries, passive ones are more stealthy and even harder to detect due to their stealthy actions. \hfill $\square$ 
 % \qed
\end{remark}

% The interactions between a distribution network and DERs can be accomplished through the SO, who is responsible for the coordination and control of DERs. To enable such synergies, private information about the DERs and DER owners is assumed to be available to the SO. Besides, the unprotected communication channels can even disclose DER owners' sensitive data in real time. 

% For example, the control of EVs' in a distribution network may require the charging status for load shifting or real-time locations depending the objectives. Therefore, the vehicle owners are compelled to send their private data to the SO, which can lead to the disclosure of sensitive data \cite{rottondi2014enabling}.  

% As a result of the sensitive data disclosure, the vehicle owners inevitably expose their real-time locations to adversaries and agonize over privacy. 

% \noindent \textbf{Assumption 1:} \textit{All participants unconditionally follow the procedures of the paradigm, i.e., each participant should never modify the input variables or the parameters, and it should calculate the outputs correctly according to the paradigm.} \hfill $\blacksquare$

% \textbf{Assumption 1} assumes that the participants in solving the optimization problem have a motivation to provide correct inputs that will lead to a reasonable result they would like to learn.% 
%
\section{Privacy Preservation Techniques}
\label{Privacy_Preservation_Techniques}

In this section, we explore the details of both mainstream and emerging privacy preservation techniques for scalable multi-agent frameworks and demonstrate their applications in DER control problems (see Tables \ref{table_DP_review}, \ref{table_ED_literature}, \ref{table_SS_literature},\ref{table_others_literature}). Furthermore, we discuss potential research and development directions for each privacy preservation technique.

\subsection{Differential Privacy} 
The concept of \emph{differential privacy (DP)}, first introduced by Dwork \cite{dwork2006calibrating,dwork2006differential}, captures the increased risk to one’s privacy incurred by participating in a database. By adding random noises to the database,  a curator (or SO) can release statistical information output of a data analysis result without compromising any individuals' privacy. Owing to its rigorous mathematical definition, DP has been a \emph{de facto} standard in developing privacy preservation/protection techniques (Note that `preservation' can imply a stronger standard than `protection'. For example, privacy is not completely preserved by 
DP since it only limits privacy leakage and does not eliminate it. In this paper, we refer to the broad concept of `protection' when using both terms). DP-based methods can quantify the privacy loss at a differential change in a database (i.e., adding or removing one entry), described by a privacy parameter $\epsilon$ that captures the privacy loss. DP ensures that privacy is preserved regardless of the combination of computations performed on the dataset, providing strong privacy guarantees against arbitrary adversaries, e.g., any re-identification attack \cite{zhao2022survey}.  
DP is a powerful privacy preservation architecture to make confidential data widely available for data analysis in the broad areas of artificial intelligence \cite{abadi2016deep, ji2014differential, wei2020federated}, power and energy systems \cite{fioretto2019differential, eibl2017differential, zhao2014achieving}, and the internet of things \cite{yin2017location, jiang2021differential}. To aid in understanding, the definition of $\epsilon$-DP is given here \cite{dwork2006differential}: 

\noindent \textbf{Definition 1}.  A randomized algorithm $\mathcal{K}$ with domain $\mathcal{C}$ is $\epsilon$-differentially private if, 
for all data sets $c_1 \in \mathcal{C}$ and $c_2 \in \mathcal{C}$ differing on at most one element, and for any possible output $S$ of the algorithm, the following inequality holds:
\begin{equation}\mathbb{P}\left[\mathcal{K}\left(c_{1}\right) = S\right] \leq e^{\epsilon} \cdot \mathbb{P}\left[\mathcal{K}\left(c_{2}\right) = S\right] \label{DP_definition}
 \end{equation}
 where $\mathbb{P}$ denotes the probability and $\epsilon$ denotes a non-negative parameter. \hfill $\blacksquare$

% for all $S \subseteq \mathcal{K}[D]$, and for and 

The parameter $\epsilon$ in \eqref{DP_definition} controls the level of privacy, i.e., a smaller $\epsilon$ implies stronger privacy guarantees, as it limits the difference in the output probabilities between adjacent datasets. 
Definition 1 provides information-theoretic protection against the maximum amount of information an adversary can acquire about any specific agent in the database, irrespective of the adversary's prior knowledge or computational capabilities. Therefore, a curator can utilize a randomized function $\mathcal{K}(\cdot)$ to mask agents' private data when releasing information. If \eqref{DP_definition} is satisfied, the released statistical information will not compromise the privacy of any individual agents. DP-based structures enjoy an \emph{ad omnia} guarantee, i.e., any information learned from the statistical database can also be obtained without directly accessing the database.

% DP has been widely applied in contingency table releases where private statistics are contained in the contingency tables. Barak \emph{et al.} in \cite{barak2007privacy} proposed a holistic solution to the problem of contingency table release, where DP, accuracy, and consistency are guaranteed simultaneously. 

 % A centralized differentially private optimal power flow mechanism was developed in \cite{dvorkin2020differentially_centralized}. However, centralized structures generically suffer from poor scalability compared to distributed or decentralized ones.

The rigorous theoretical foundation of DP has led to extensive privacy preservation applications in multi-agent systems and DER control in power systems. 
% \revsecond{{\bf{REV 2.7}}}{\color{red}
For example, DP-based methods can add calibrated noise into smart meter data and/or into the computing process, therefore protecting the attributes of any single individual's smart meter readings without affecting grid-level and/or DER-level objectives. DP-based methods have been integrated into scalable methods to achieve privacy protection for power system applications, such as protecting consumers' smart meter data \cite{Gough9411689}, power flow problems \cite{dvorkin2020differentially,ryu2021privacy}, and EV charging control \cite{lee2024multilevel}. 
% PGD \cite{hale2017cloud}, AvegC \cite{fiore2019resilient}, and MARL \cite{lee2024multilevel}. 
% }
% 
% This ensures that personal data remains secure without affecting grid-level and/or DER-level objectives.
% By adding well-calibrated noises into the \replaced{computing}{computation} process, DP-based methods can obscure the attributes of any single individual's data (e.g., smart meter readings) without affecting grid-level and/or DER-level objectives. 
% 
Based on DP, Hale and Egerstedt \cite{hale2017cloud} develop a privacy-preserving primal-dual  optimization framework for multi-agent convex programs and solve it using the PGD. It keeps each agent's state trajectory private from all other agents and any external eavesdroppers. Han \emph{et al.} \cite{han2016differentially} develop a distributed privacy-preserving optimization algorithm based on DP to preserve the privacy of the participating agents in constrained optimizations. To broaden the range of adversaries, Fiore and Russo \cite{fiore2019resilient} design a DP-based consensus algorithm for multi-agent systems where a subset of agents could be honest-but-curious. 
In \cite{Gough9411689}, a DP-based privacy preservation algorithm is developed to protect consumers’ smart meter data. Dvorkin \emph{et al.}  \cite{dvorkin2020differentially} develop an adversarial inference model based on DP that first questions the privacy properties of distributed OPF. Subsequently, the authors develop a differentially private variant of the ADMM to ensure information privacy during information exchanges between neighbors. This model is later extended in \cite{ryu2021privacy} for the distributed optimization of AC power flow problems. In \cite{dong2018privacy}, a DP-based aggregation algorithm is proposed to compensate for solar power fluctuations and protect customers' personal information. In \cite{fioretto2019differential}, a DP-based obfuscation mechanism is proposed with guarantees of AC feasibility to protect the private parameters of transmission lines and transformers. 
% 
% \revsecond{{\bf{REV 2.7}}}{\color{red} 
Lee and Choi 
\cite{lee2024multilevel} develop a DP-based multilevel deep RL algorithm for privacy-preserving EV charging operations, ensuring optimized data privacy, revenue, and energy costs. 
% }

% \hspace{0mm}\revsecond{{\bf{REV 2.7}}}{\color{red}
To summarize, DP has become a foundational principle in the privacy protection field, with wide-ranging power system applications, and is evolving rapidly alongside advancements in developing scalable multi-agent frameworks.
% }
% 
The potential of DP can be further explored from the following directions: (1) \emph{Reduce the privacy-accuracy gap}. DP-based methods commonly suffer from loss of accuracy caused by the added noise. Research shows that a balance between privacy and accuracy can be achieved via the design of carefully calibrated noises \cite{smith2011privacy,chaudhuri2011differentially}. (2) \emph{Extension of DP for both privacy (passive adversaries) and security (active adversaries) scenarios}. When faulty agents maliciously deviate from the computing policy or network communication protocol, the effectiveness of DP in maintaining both privacy and security can be compromised \cite{miao2022linear,ye2019differentially,zhang2023evolution}. The co-design of a privacy-preserving and cybersecure multi-agent framework is worth further investigation. (3) \emph{Enhanced compatibility for the next generation of learning-aided methods}. DP has shown strong  cohesion in preserving privacy for learning-aided methods, including training train neural networks for deep learning models \cite{bu2020deep}, employing stochastic gradient descent for machine learning \cite{wu2020value}, reducing sample complexity with new expansion 
on DP \cite{pillutla2024unleashing}. The rapid evolution of DP also shows strong compatibility in addressing emerging privacy concerns in learning-aided approaches, such as the deployment of large language models in the electric power sector \cite{majumder2024exploring}.

% {\color{blue}
\begin{remark} 
% \normalfont
% \revfirst{{\bf{REV 1.9}}}
% 
The DP technique can be categorized into global DP and local DP depending on whether the server (e.g., SO/coordinator/aggregator) can be trusted during the computing process. In global DP, privacy is maintained by adding noise at the central server level (e.g., database query), while in local DP, privacy is preserved by individuals before sending any raw data to the central server (e.g., adding noise to smart meter data). 
In practice, choosing between global and local DP depends on the use case and the level of trust in the central server, i.e., global DP is common when a central server is trusted \cite{hale2017cloud}, while local DP is preferred when trust is an issue and individual-level privacy must be ensured \cite{fiore2019resilient}. 
Note that, the definition of centralized, distributed, and decentralized information exchange structures are categorized based on their communication patterns, i.e., how private information flows between different parties, while the categorization of DP reflects the trustworthiness of the central server.  \hfill $\square$ 
\end{remark}
% }

% \newpage

\scriptsize
%\begin{longtable}{m{3em} m{2em} m{6em} m{9.5em} m{18em}} 
\begin{longtable}{ m{2.7em}  m{1.7em}  m{6em} m{5.3em} m{7.8em} m{14.1em}} 
\caption{Differential-Privacy (DP)-based Scalable and Privacy-Preserving Methods.} \label{table_DP_review} \\
    \toprule
Method & Ref. & Problem  & Structure  & Adversaries & Key Features\\
    \midrule
    \endfirsthead
    \caption[]{(Continued)}\\
    \toprule
Method & Ref. & Problem  & Structure  & Adversaries & Key Features\\
    \midrule
    \endhead
    \bottomrule
    \endfoot

    % \sisetup{table-format=2.4}% table-column-width =2.5cm
%     \centering \footnotesize
%     \begin{tabular}{ m{3em}  m{4em}  m{9em} m{7em} m{9em} m{19em}} 
%         \toprule
% Method & Reference & Problem  & Structure  & Adversaries & Key Features\\
%     \midrule

DP  & \cite{hale2017cloud}  &  Multi-agent convex programs & Decentralized & Honest-but-curious agents, the cloud, ($\epsilon,\delta$)-DP &  1-Require a trusted cloud computer; 2-the cloud adds noise to data. \\
    
DP & \cite{fiore2019resilient}   & Consensus for multi-agent system  & Distributed & Byzantine and malicious agents, $\epsilon$-DP & 1-A subset of agents is adversarial; 2-achieve resilient asymptotic consensus with correctness, accuracy and DP properties. \\

DP & \cite{huang2012differentially}   & Consensus for multi-agent system  & Distributed & Honest-but-curious agents & 1-Server-based  randomized mechanism; 2-adversaries can observe the messages and states of the server and a subset of the clients. \\

DP & \cite{nozari2016differentially} &  Convex constrained optimization  & Distributed & Honest-but-curious agents & 1-Individual objective function is kept private; 2-both input and output-perturbation methods. \\

DP & \cite{wang2022differentially} &  Stochastic aggregative games  & Distributed & Honest-but-curious agents, eavesdroppers & 1-Seek Nash equilibrium in stochastic aggregative games; 2-retain smoothness and regularity properties. \\

DP & \cite{dvorkin2020differentially} &  Optimal power flow  & Distributed & An adversarial inference
model & 1-Develop an adversarial inference
model for OPF; 2-introduce static and dynamic random perturbations of OPF sub-problem; 3- $\epsilon$-DP. \\

DP & \cite{ryu2021privacy} &  Optimal power flow  & Distributed & A hypothetically strong adversary   & 1-DP projected subgradient; 2-non-differentiable concave objective function. \\

DP & \cite{han2016differentially} &  Resource allocation problems   & Distributed & Adversaries and their collaboration with some users  & 1-The privacy guarantee is proved using the adaptive composition theorem; 2-view the differentially private algorithm as stochastic gradient descent; 3-implementation for EV charging control. \\

DP & \cite{wang2024robust} &  Multi-agent systems & Distributed & Passive inference adversaries \cite{zhu2019deep,melis2019exploiting}  & 1-Constrained consensus that can ensure both accurate convergence and $\epsilon$-DP; 2-without requiring the Lagrangian function to be strictly convex/concave. \\

DP & \cite{nozari2017differentially} &  Average consensus  & Distributed & Honest-but-curious agents, eavesdroppers  & 1-Establish the impossibility of exact average for differentially private algorithms; 2-design a linear consensus algorithm with unbiased consensus value. \\

DP & \cite{dvorkin2022privacy} &  Convex optimization programs  & Centralized & $\epsilon$-DP  & 1-Express the optimization variables as functions of the random perturbation;  2-employ chance-constrained linear decision rule optimization to impose feasibility requirement. \\

DP & \cite{munoz2021private} &  Resource allocation problems & Centralized & $(\epsilon, \delta)$-DP & 1-Solve linearly-constrained optimization problems with hard requirement on constraint violations; 2-truncated Laplace mechanism. \\

DP & \cite{lee2024multilevel} &  EV charging control with solar PVs and ESSs & Centralized training with decentralized execution & $(\epsilon, \delta)$-DP & 1-Multi-level deep RL structure for DERs; 2-agents cooperate to maximize the revenue of smart charging station. \\

DP & \cite{huang2024differential} &  Optimization with gradient tracking & Distributed & $\epsilon$-DP & 1-Add noises to the decision variables and the estimate of the aggregated gradient; 2-prove the impossibility of simultaneous exact convergence and DP preserving. \\

DP & \cite{atmaca2024privacy} & Queries of charging stations for EVs & Centralized & $(\epsilon, \delta)$-geo-indistinguishable, (honest-but-curious service providers, cloud architecture)  & 1-EVs obfuscate their query locations; 2-use approximate geo-indistinguishability as a generalization of local DP. \\

DP & \cite{wang2023tailoring} &  Multi-agent systems & Distributed & $\epsilon$-DP & 1-Tailor gradient methods for differentially private distributed optimization; 2-based on static and dynamic consensus gradient methods. \\

DP & \cite{zhao2018privacy} &  Distributed energy management  & Distributed & $(\epsilon, \delta)$-DP, out-neighbors, eavesdroppers & 1-A secret-function-based privacy-preserving algorithm; 2-nodes add zero-sum and exponentially decaying noises to the original data for communications. \\

\bottomrule

\end{longtable}
\vspace{-0.4cm}
\scriptsize {\noindent Ref.: Reference ~~ DP: Differential privacy}
\normalsize
 
% DP & \cite{} &  Multi-agent systems & Distributed &  & 1-; 2-. \\
%     \bottomrule
%     \end{tabular}
%     \vspace{2mm}

% \raggedright
% \footnotesize{$~~~~~~~~~~$ DP: Differential privacy}
% \end{table*}  

\subsection{Cryptographic Methods}

The protection of privacy in multi-agent frameworks can also be achieved through the integration of cryptographic techniques. A typical cryptosystem involves encryption and decryption operations, which can be integrated into distributed or decentralized information exchange structures to protect private information while ensuring scalable computing. This paper focuses on \textit{encryption-decryption-based} and \textit{secret sharing-based} methods.

\subsubsection{Encryption-decryption-based methods} 

Encryption-decryption (ED)-based methods utilize a cryptosystem that typically consists of three components: An encryption algorithm, a decryption algorithm, and key management.
Specifically, a plaintext $m$ is encrypted into a ciphertext $\mathcal{E}(m)$ using an encryption function $\mathcal{E}(\cdot)$. By applying a decryption function $\mathcal{D}(\cdot)$ to the ciphertext, the original plaintext can be correctly retrieved as $m = \mathcal{D}(\mathcal{E}(m))$. 
\begin{figure}[!htb]
% \vspace*{-2mm}
    \centering
    \includegraphics[width=0.7\textwidth, trim = 0mm 0mm 0mm 0mm, clip]{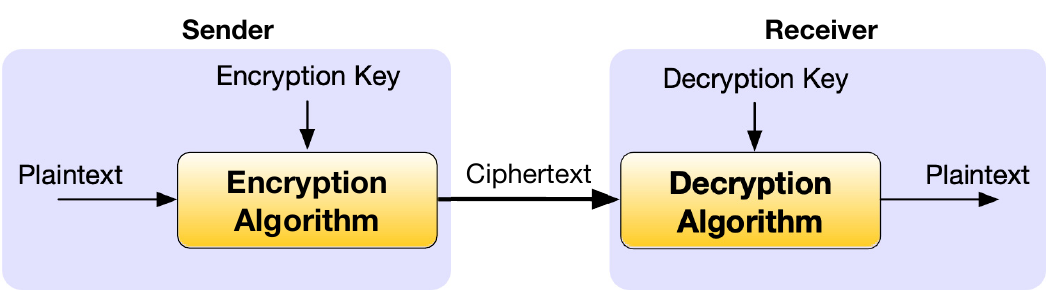}
    % \vspace*{-3mm}
    \caption{Secure communications between a sender and a receiver using a cryptosystem.}
    % \vspace*{-4mm}
    \label{Cryptosystem}
\end{figure}
Fig. \ref{Cryptosystem} shows the realization of secure communications using a  cryptosystem. A sender sends some sensitive plaintexts to a receiver in the form of ciphertexts using a cryptosystem such that any party intercepting/eavesdropping on the communication channel only has access to the ciphertexts, instead of knowing the plaintexts.

Among various cryptosystems, homomorphic cryptosystems are well-suited for multi-agent computing and communications. Essentially, a homomorphic cryptosystem enables users to perform computations on encrypted data without having to decrypt it first. The homomorphic properties are typically necessary for performing secure arithmetic operations in multi-agent systems \cite{lu2018privacy,huo2021encrypted, zhang2018enabling, hadjicostis2020privacy, wu2021privacy, yuan2023fully}, 
% \revsecond{{\bf{REV 2.7}}}{\color{red} 
offering significant potential for power system applications, such as optimal power flows \cite{wu2021privacy}, DER management \cite{cheng2021homomorphic,wang2023enhancing}, and economic dispatch \cite{chen2024quantized,mu2024decentralized}.
% } 
%
Homomorphic schemes can be classified according to the types of mathematical operations that can be performed on ciphertexts: 1) \emph{Partially homomorphic} that supports either addition or multiplication operation, but not both simultaneously, and 2) \emph{fully homomorphic} that concurrently support addition and multiplication operations. For a cryptosystem to be fully homomorphic, it needs to satisfy:
\begin{subequations} \label{fully_homorphic}
\begin{align}
\mathcal{D}(\sum_{e=1}^{\bar{e}} \mathcal{E}(m_{e})) &=\sum_{e=1}^{\bar{e}} m_{e} \label{3a} \\
\mathcal{D}(\prod_{e=1}^{\bar{e}} \mathcal{E}(m_{e}))&=\prod_{e=1}^{\bar{e}} m_{e} \label{3b} 
\end{align}
\end{subequations}
where  $m_{e}$ denotes the $e$th plaintext and $\bar{e}$ denotes the total number of plaintexts.

The Paillier cryptosystem \cite{paillier1999public}, for example, is partially homomorphic, allowing the addition of two ciphertexts and only the multiplication of a ciphertext by a plaintext. The Paillier cryptosystem is constructed by generating a set of public and private keys, where plaintexts are encrypted using the public key, and ciphertexts can be decrypted using the private key.  The security of the Paillier cryptosystem is based on the computational complexity of the decisional composite residuosity assumption (DCRA) \cite{paillier1999public}. 
In specific, the hardness of the DCRA comes from the fact that for large composite numbers, the problem of distinguishing residues is computationally infeasible. The difficulty is similar to the hardness of factoring large composite numbers. The spectrum of adversaries in ED-based strategies can be proven from the secure multi-party computing perspective against different adversaries, such as honest-but-curious agents, external adversaries, and the SO/coordinator/aggregator \cite{li2010secure,zhang2018enabling,zhang2018admm,wu2021privacy,huo_control_letter,hadjicostis2020privacy}.

% The Paillier cryptosystem has been widely adopted in achieving PP for solving constrained optimization problems \cite{zhang2018enabling,zhang2018admm,hadjicostis2020privacy,huo_control_letter}. 
% \revsecond{{\bf{REV 2.2}}}\noindent {\color{red}
\noindent \textbf{Example 1} (Integration of ED into PGD).
Consider the PGD in Section \ref{PGD}, we give an example of integrating ED into PGD-based scalable multi-agent frameworks. % 
\begin{figure}[!htb]
% \vspace*{-2mm}
    \centering
    \includegraphics[width=0.6\textwidth, trim = 0mm 0mm 0mm 0mm, clip]{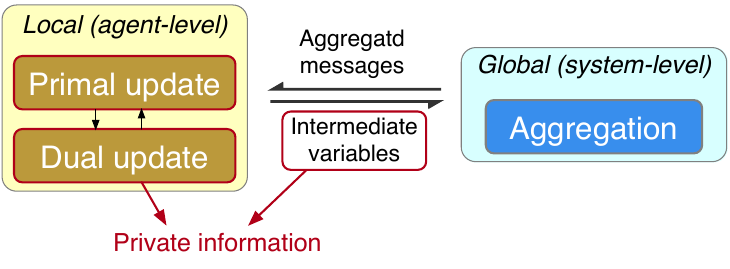}
    % \vspace*{-3mm}
    \caption{Private information is calculated and transmitted in primal-dual-based computing schemes \cite{huo2021encrypted}.}
    % \vspace*{-4mm}
    \label{primal_dual_architecture}
\end{figure}
As shown in Fig. \ref{primal_dual_architecture}, plaintexts are calculated and transmitted directly without privacy protection between agents and the system (coordinator) in a primal-dual-based computing scheme.
By using ED, agents can encrypt private information (e.g., decision variables, private coefficients, subgradients, objective functions \cite{lu2018privacy,huo2021encrypted}) and then communicate with the coordinator only via ciphertexts. The coordinator can access, aggregate, and compute ciphertexts based on the cryptosystem's homomorphic properties. The primal and dual updates can be executed in the space of ciphertexts. \hfill $\square$ 
% }

% Fig. \ref{primal_dual_architecture} shows a possible privacy-preserving computing architecture of PDG with the Paillier cryptosystem. 

% Herein, we present the key features of the Paillier cryptosystem, aiming at exampling (ED-based) strategies. In Paillier cryptosystem, a manager first  generates two sets of keys, i.e., the public key $(p,g)$ and the private key $(\eta,\mu)$. The public key can be obtained by $p=vw$, $g = p+1$ where $v$ and $w$ are large prime numbers of the same length. After generating the public key, the private key $(\eta,\mu)$ can be calculated via  $\eta = \phi(p)$ and $\mu = \phi(p)^{-1} \bmod p$ where $\phi(p) = (v-1)(w-1)$ and $\bmod$ denotes the modulo operation. Note that $v$ and $w$ can also have different lengths with a more complex private key selection procedure, which can be referred to \cite{paillier1999public}.
% Having the keys ready, the manager publicizes the public key while keeping the private key to itself. The ciphertext can be decrypted with the private key by
% \begin{equation}
%  \tilde{m}_z =  \left\lfloor\frac{\left(c^{\eta} \bmod p^{2}\right)-1}{p}\right\rfloor \mu \bmod p,
%  \label{2}
% \end{equation}
% where $\tilde{m}_z$ denotes the decrypted message and $\lfloor \cdot \rfloor$ denotes the floor of a real number. 

% Zhang \emph{et al.} in \cite{zhang2018admm} integrated Paillier cryptosystem into decentralized optimization to achieve PP where the decentralized algorithm was designed first without PP based on alternating direction method of multipliers. 

ED-based methods have been widely integrated within the design of scalable and privacy-preserving multi-agent frameworks. Lu and Zhu \cite{lu2018privacy} develop homomorphic-encryption-based schemes that can achieve secure multi-party computing with  privacy preservation guarantees. Along this research direction, a privacy-preserving decentralized multi-agent cooperative optimization paradigm is proposed in \cite{huo2021encrypted} by integrating additively homomorphic cryptosystem into decentralized optimization. In \cite{zhang2018enabling}, a decentralized privacy-preserving algorithm based on the Paillier cryptosystem is developed to protect agents' intermediate variables in distributed systems. Hadjicostis \emph{et al.} \cite{hadjicostis2020privacy} develop a privacy-preserving AvgC method using the Paillier cryptosystem. It allows agents to reach a consensus on the average of their initial integer values while maintaining the confidentiality of these values in the presence of honest-but-curious agents.

% \hspace{0mm}\revsecond{{\bf{REV 2.7}}}{\color{red} 
In the power systems field, ED-based methods are well-suited for real-world implementation due to their compatibility with complex computing and communication structures. This compatibility enables the secure transmission of sensitive information, including customer load profiles, power system operational status, and operator control commands.
Moreover, the scalability and privacy protection capabilities of ED-based methods enhance the industry relevance of multi-agent frameworks for DER management, aligning them with electrical engineering standards such as IEC 62351, IEEE 1815-2012 (DNP3), and NERC reliability standards \cite{iec_62351_standards,DNP3_standards,NERC_reliability_standards}.
% } 
To preserve the private voltage and current measurements, Wu \emph{et al.} \cite{wu2021privacy} develop a privacy-preserving distributed OPF algorithm based on partially homomorphic cryptosystems. To eliminate the privacy concerns of economic dispatch problems in microgrids, a homomorphically encrypted algorithm is developed to achieve consensus without disclosing agents' private or sensitive state information \cite{chen2022privacy}. He \emph{et al.} \cite{he2017efficient} develop a computationally efficient data aggregation scheme based on public key cryptography to  prevent the extraction of consumers' electricity consumption information against internal and external attackers.

ED-based methods continue to evolve as one of the mainstream privacy preservation measures, attracting significant attention for secure computing in multi-agent systems. Here are some future directions for ED-based methods:  (1) \emph{Decrease the computing overhead}. The complexity of a cryptosystem, the key length, and the size of encrypted or decrypted data all largely impact the computing cost. Designing computationally efficient cryptographic algorithms is critical for enabling scalable and privacy-preserving DER operations \cite{he2017efficient}.  (2) \emph{Trustworthy key management}. In establishing and executing cryptographic protocols, participants must manage keys (initialize, update, rotate, or revoke) in a secure way. The leakage of keys can lead to direct corruption of a cryptographic scheme. Therefore, establishing trustworthy key management is essential for controlling DERs with tremendous end-users. (3) \emph{Interoperability within industrial standards}. Cryptographic algorithms should be deployed in an interoperable way with modern electric engineering standards. There is also the need for standardized cryptographic practices that can be uniformly applied across various customers and vendors in the electric power sector.

\scriptsize
%\begin{longtable}{m{3em} m{2em} m{6em} m{9.5em} m{18em}} 
\begin{longtable}{ m{2.7em}  m{1.7em}  m{6em} m{5.3em} m{7.8em} m{14.1em}} 
%\begin{longtable}{ m{2.8em}  m{1.8em}  m{6em} m{5.3em} m{9em} m{16em}} 
\caption{Encryption-Decryption (ED)-based Scalable and Privacy-Preserving Methods.} \label{table_ED_literature} \\
    \toprule
Method & Ref. & Problem  & Structure  & Adversaries & Key Features\\
    \midrule
    \endfirsthead
    \caption[]{(Continued)}\\
    \toprule
Method & Ref. & Problem  & Structure  & Adversaries & Key Features\\
    \midrule
    \endhead
    \bottomrule
    \endfoot

ED  & \cite{lu2018privacy}  & Projected gradient-based algorithm  &  Distributed & Honest-but-curious agents, eavesdroppers, the system operator & 1-Based on secure multiparty computation; 2-develop private and public key secure computation algorithms.\\
    
ED  & \cite{huo2021encrypted}  & Multi-agent cooperative optimization  &  Decentralized & Honest-but-curious agents, eavesdroppers, the system operator & 1-Applicable on general primal-dual-based algorithms; 2-real-world experimental demonstration.\\

ED  & \cite{zhang2018enabling}  & Constrained decentralized optimization  &  Decentralized & Honest-but-curious agents, eavesdroppers & 1-Integrate partially homomorphic cryptography; 2-applicable to average consensus problem.\\

ED  & \cite{hadjicostis2020privacy}  & Average consensus  & Distributed & Honest-but-curious agents & 1-Assume the presence of a trusted node; 2-privacy preservation via multiple encrypted ratio consensus iterations.\\

ED  & \cite{wu2021privacy}  & Optimal power flow  & Distributed & Honest-but-curious agents, external eavesdropper, the system operator & 1-ADMM-based structure; 2-encrypt the dual update by the Paillier cryptosystem; 3-relax the augmented term of the primal update.\\

ED  &  \cite{he2017efficient} & Smart meter data aggregation  & Decentralized & External and internal adversaries & 1-Boneh-Goh-Nissim public key cryptography; 2-consider both privacy, authentication, and integrity; 3-involve a trusted third party and an aggregator.\\

ED & \cite{mu2024decentralized} &  Optimal dispatch of wind farms and shared ESSs & Decentralized & Other wind farms (Honest-but-curious)  & 1-Wind power uncertainty is handled through chance constraints; 2-include physical and virtual ESS components. \\

ED & \cite{chen2024quantized} & Distributed economic dispatch of microgrids & Distributed & Honest-but-curious nodes, eavesdroppers & 1-Coordinate the power outputs of distributed generators; 2-based on Paillier cryptosystem; 3-converge to the optimal solution under finite quantization levels. \\

ED & \cite{hu2023privacy} &  IoT-based active distribution network  & Distributed & Eavesdroppers & 1-Homomorphically encrypted energy management system for economic coordination and power sharing; 2-preserve privacy of distributed generators and customers' loads. \\

ED & \cite{wang2023enhancing} &  Vehicle-to-vehicle energy trading & Distributed & Sybil attack, double spending, DoS  & 1-Propose an EV leader election based on cryptography that can secure transfer of energy and value; 2-adopt the sharding technique to enhance the system’s scalability. \\

ED & \cite{ma2023novel} &  Energy trading in microgrids & Centralized & Honest-but-curious adversary, false injection attack$^\star$, message falsification attack$^\star$  & 1-Secure and privacy-preserving energy trading under the untrusted server in the microgrid; 2-evaluate the trading price and power flow via homomorphic energy data evaluations. \\

ED & \cite{cheng2021homomorphic} &  Distributed energy management system & Distributed & Eavesdroppers, man-in-the-middle attack$^\star$  & 1-Bus-level agent-based distributed primal-dual subgradient algorithm; 2-fully homomorphic encryption. \\

ED & \cite{froelicher2020scalable} &  Distributed learning & Distributed & Up to $N-1$ colluding parties  & 1-Enable the privacy-preserving execution of the cooperative gradient descent; 2-build on a multi-party fully homomorphic encryption scheme. \\

% ED & \cite{mohammadali2021privacy} &  Metering data aggregation & Distributed & semi-honest adversary  & 1-; 2-. \\

ED & \cite{shoukry2016privacy} &  Quadratic optimization problem & Distributed & Semi-honest colluding parties (agents coalitions, cloud coalitions, target node coalitions)  & 1-Protect privacy-sensitive objective function and constraints; 2-privacy guarantees are analyzed using  zero-knowledge proof. \\

% ED & \cite{} &  Multi-agent systems & Distributed &  & 1-; 2-. \\

% ED & \cite{} &  Multi-agent systems & Distributed &  & 1-; 2-. \\

% ED & \cite{} &  Multi-agent systems & Distributed &  & 1-; 2-. \\

% DP & \cite{} &  Multi-agent systems & Distributed &  & 1-; 2-. \\
\bottomrule

\end{longtable}
\vspace{-0.4cm}
\scriptsize{\noindent $^\star$Refers to active adversaries ~~ Ref.: Reference ~~ ED: Encryption-decryption}
\normalsize

\subsubsection{Secret sharing-based methods}

Secret sharing (SS) is a lightweight cryptographic protocol that can split a secret into multiple shares and distribute the shares among a group of participants. The essential idea behind SS is to ensure that the secret can only be reconstructed by combining an adequate number of shares. Meanwhile, any subset of shares smaller than a threshold yields no useful information about the secret. 
Shamir’s SS \cite{shamir1979share} is a well-known SS scheme in which the secret shares are generated using a polynomial. Specifically, Shamir’s SS is developed based on the concept of polynomial interpolation, defined as \cite{humpherys2020foundations}: 

\noindent \textbf{Theorem 1} (\textit{Polynomial interpolation}). Let $\{(z_{1}, f_{1}), \ldots,$ $(z_{\bar{d}}, f_{\bar{d}})\}$ $\subseteq \mathbb{R}^{2}$ be a set of points whose values of $z_d$ are all distinct. Then, there exists a unique polynomial $f^{(\bar{d}-1)}$ of degree $\bar{d}-1$ that satisfies $f_d = f^{(\bar{d}-1)}(z_d), \forall d=1,\ldots,\bar{d}$.  \hfill $\blacksquare$

Theorem 1 states that a minimum number of $d+1$ points equal to the degree of the polynomial are required to reconstruct the secret. This ensures information-theoretic security, meaning that even if an adversary obtains some shares, it is impossible to reconstruct the secret unless they have acquired the \emph{quorum} number of shares.

\begin{figure}[!htb]
% \vspace*{-2mm}
    \centering
    \includegraphics[width=0.7\textwidth, trim = 0mm 0mm 0mm 0mm, clip]{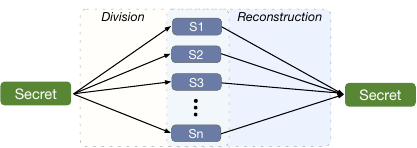}
    % \vspace*{-3mm}
    \caption{An illustration of Shamir's secret sharing, showing the division of shares and the secret reconstruction process.}
    % \vspace*{-4mm}
    \label{Fig_Shamir_secret_sharing}
\end{figure}

% \revsecond{{\bf{REV 2.2}}}\noindent {\color{red}
\noindent \textbf{Example 2} (Division of shares and secret reconstruction in SS).
The procedures of Shamir’s SS \cite{shamir1979share}, including the division of shares and the reconstruction of secrets, are given in Fig. \ref{Fig_Shamir_secret_sharing}. 
Shamir’s SS can be executed in three steps: 
\textit{(1) Polynomial generation:} A manager (secret holder) constructs a random polynomial 
$f(z) = s + c_1z + \cdots + c_{\bar{d}-1}z^{\bar{d}-1}$,
where $s$ denotes the secret, the coefficients $c_1,\ldots,c_{k-1}$ are randomly chosen from a uniform distribution in an integer field
$\mathbb{E} \triangleq [0,e)$, where $e$ denotes a large prime number;
\textit{(2) Division of shares:} The manager computes the shares with a non-zero integer input and obtains the output, e.g., set $i=1,\ldots,n$ to retrieve S$_i$ $= (i, f(i) ~ (\bmod ~e))$. Then, it distributes the share S$_i$ to the $i$th agent;
% , where $f_i = f(i) \bmod e \label{f_i}$; 
% 
\textit{(3) Secret reconstruction:} Therefore, based on Theorem 1, at least $\bar{d}$ points are needed to reconstruct the polynomial and calculate the secret $s$. 
The efficient division of shares and reconstruction of secrets make SS highly suitable for privacy-preserving applications in multi-agent systems, including cryptographic key distribution, access control management in distributed systems \cite{beimel1996secure}, and the protection of sensitive data \cite{zhang2018cloud,li2019privacy}.
SS has also shown potential in protecting privacy for  controlling grid-edge energy resources for power system applications, including electricity theft detection  \cite{nabil2019ppetd} and vehicle-to-grid integration \cite{rottondi2014enabling,huo2022distributed}.
% } 
% goo
Adopting SS, Nabil \emph{et al.} \cite{nabil2019ppetd} design a privacy-preserving detection scheme to identify electricity theft from malicious consumers. Only masked meter readings from consumers are collected and sent to the SO, ensuring privacy protection and preventing data leakage. In \cite{huo2022distributed}, an SS-based cooperative EV charging control protocol is developed to achieve overnight valley filling without compromising the privacy of EV owners' charging profiles.  The distributed protocol enjoys high computation efficiency and accuracy. In \cite{rottondi2014enabling}, a privacy-preserving communication protocol based on SS is proposed for vehicle-to-grid integration. The proposed protocol ensures that the existing battery charge level, the quantity of replenished energy, and the duration of EVs being plugged in remain undisclosed to aggregators.

In summary, as a threshold scheme based on polynomials and finite geometries, Shamir's SS is well-suited for secure computation and key sharing in cryptographic applications among multiple stakeholders. Some potential future directions for SS-based methods: (1) \emph{Homomorphic secret sharing}. 
Combining SS with homomorphic properties allows computations to be performed on the shared data without revealing the secret. It is worth investigating efficient homomorphic operations within the SS framework for privacy-preserving computations in  scalable multi-agent systems. (2) \emph{Threshold cryptography in dynamic environments}. Traditional SS requires a predefined number of participants. However, real-world power system applications often involve dynamic groups and changing environments. Research could focus on adapting SS to handle dynamic groups, where agents can join or leave without compromising the shared secret. 
% 
% \revsecond{{\bf{REV 2.7}}}{\color{red}
(3) \emph{Power system applications}. Future research could focus on how SS can be adapted in different power system communication topologies, such as federated, distributed, and decentralized mechanisms, to protect sensitive DER data. 
% line of applications, such as consensus mechanisms, blockchain systems, and distributed storage systems.
% }

\scriptsize
%\begin{longtable}{m{3em} m{2em} m{6em} m{9.5em} m{18em}} 
\begin{longtable}{ m{2.7em}  m{1.7em}  m{6em} m{5.3em} m{7.8em} m{14.1em}} 
\caption{Secret Sharing (SS)-based Scalable and Privacy-Preserving Methods.}
    \label{table_SS_literature} \\
    \toprule
Method & Ref. & Problem  & Structure  & Adversaries & Key Features\\
    \midrule
    \endfirsthead
    \caption[]{(Continued)}\\
    \toprule
Method & Ref. & Problem  & Structure  & Adversaries & Key Features\\
    \midrule
    \endhead
    \bottomrule
    \endfoot
    
    SS  & \cite{li2019privacy} & Average consensus & Distributed & Honest-but-curious agents & 1-Achieve  security in clique-based networks; 2-allow weaker model of active attacks.\\
    
    SS  & \cite{huo2022distributed} & Multi-agent cooperative optimization & Distributed &  Honest-but-curious agents, eavesdroppers & 1-Coordinate EVs to achieve overnight valley filling; 2-applicable to projected gradient-based algorithms.\\

    SS & \cite{zhang2020consensus} & Average consensus & Distributed & Honest-but-curious agents, eavesdroppers & 1-Agents reach an agreement without exposing their individual states until the agreement is reached; 2-resistant to the collusion of any given number of neighbors.\\ 

    SS & \cite{tian2022secret} & Multi-party collaborative optimization & Distributed & Honest-but-curious agents & 1-Exchange shares between agents; 2-decomposition and coordination among agents for convergence.\\ 

    SS  & \cite{laufer2024privacy} & Partitioned DER control  & Decentralized &  Server (the resource operator), other agents  & 1-Ensure client privacy and system integrity; 2-applicable to run on resource constrained embedded systems.\\

    SS  & \cite{wagh2023distributed} & Smart metering data aggregation  & Distributed & Honest-but-curious smart meters and service providers,  dishonest majority of aggregators$^\star$   & 1-Can verify the integrity of the spatio-temporal metering data; 2-consider a malicious adversarial model with a dishonest majority of aggregators.\\

    SS  & \cite{bonawitz2017practical} & Federated learning  & Decentralized & Honest-but-curious, active adversary$^\star$   & 1-Communication-efficient and failure-robust protocol for secure aggregation of high-dimensional data; 2-maintain security even if a subset of users drop out at any time.\\

    SS  & \cite{rottondi2014enabling} &  Vehicle-to-grid communication infrastructure  & Distributed & Honest-but-curious (aggregator, collusion of aggregators, anonymizer)    & 1-Schedule EV charge/discharge times; 2-protect users’ traveling habits, the current battery level, and the amount of refilled energy.\\

    SS  & \cite{huo_TCNS_SS_2024} &  DER aggregation and control & Hierarchical & Honest-but-curious agents, external eavesdroppers    & 1-Develop a hierarchical DER aggregation and control framework; 2-privacy-preserving optimization based on SS with privacy protection guarantees.\\

    % SS  & \cite{rottondi2014enabling} &   & Decentralized &    & 1-; 2-.\\
    
    \bottomrule
    
\end{longtable}
\vspace{-0.4cm}
\scriptsize{\noindent $^\star$Refers to active adversaries ~~ Ref.:Reference ~~ SS: Secret sharing}
\normalsize

\subsection{Other miscellaneous and emerging methods}

\subsubsection{State decomposition} 

Wang \cite{wang2019privacy} initiates the concept of state decomposition (SD) that can achieve AvgC while protecting the privacy of all participating agents.  In SD, an agent decomposes its state into two distinct substrates, with only one substrate visible to others, thus protecting the actual value of the original state. In contrast to DP-based methods that rely on adding additional noises, SD ensures convergence of the AvgC to the desired value without any accuracy error.  The authors also extend SD on a dynamic consensus algorithm of multi-agent systems and apply it to the formation control of multiple mobile robots \cite{zhang2022privacy_wang}. 

Following this line of research, Wang \emph{et al.} \cite{wang2021privacy} design an SD-based privacy-preserving consensus algorithm where each agent is decomposed into homologous subagents based on the number of its neighbors. The homologous subagents exchange information directly, while 
the information interaction between non-homologous subagents  is encrypted by homomorphic cryptography. In \cite{zhang2022privacy}, an SD mean-subsequence-reduce algorithm is designed to address privacy preservation in the resilient consensus of discrete-time multi-agent systems. The designed method considers the worst-case malicious behaviors against active adversarial agents who may update their state values in a completely arbitrary way. To summarize, SD-based approaches can effectively eliminate numerical errors caused by the accuracy-privacy trade-off. The research on SD requires continued efforts to generalize this approach to scalable multi-agent computing frameworks for DER control problems.

\subsubsection{Noise injection}

Analogous to DP, noise injection (NI) or perturbation-based methods add random noises/offsets to the private data to ensure privacy-preserving computing and communications \cite{manitara2013privacy,mo2016privacy,charalambous2019privacy,mao2020privacy,huo2023privacy}. Typical injected noises include independent and exponentially decaying Laplacian noise \cite{li2020privacy}, Gaussian noise \cite{mo2016privacy,zou2024optimal,charalambous2019privacy,huo2023privacy}, and certain conditional noises \cite{mao2020privacy}. 

Apart from privacy, accuracy and algorithm efficiency are two important attributes that are often considered in designing NI-based methods. In \cite{li2020privacy}, a subspace perturbation method is developed to achieve privacy-preserving distributed optimization with a focus on circumventing the privacy and accuracy trade-off. By adding and subtracting random noises to the consensus process, Mo and Murray \cite{mo2016privacy} develop a privacy-preserving AvgC algorithm to guarantee the privacy of the initial state while achieving exact consensus. 
Charalambous \emph{et al.} \cite{charalambous2019privacy} design a privacy-preserving ratio consensus algorithm that can converge to the exact average of the nodes’ initial values, even in the presence of bounded time-varying delays. In \cite{zou2024optimal}, a privacy-preserving transmission scheduling strategy is proposed to defend against eavesdropping, which demonstrates the correlation between the optimal transmission decision and the intensity of the injected noise. 
To summarize, NI-based methods use noise addition similar to DP-based approaches, but they aim more at overcoming the algorithm efficiency limitations and lifting the privacy-accuracy trade-offs.
Notably, existing NI-based structures have demonstrated the ability to reduce computing and communication overhead. Future research could investigate how varied NI methods can further improve algorithm performance to a new level.

\subsubsection{Garbled circuit}

Hardware-based methods such as Boolean/arithmetic circuits can both be utilized to achieve secure computation between multiple parties \cite{ben2008fairplaymp}. The classic garbled circuit (GC) is initially proposed by Yao in \cite{yao1982protocols} to address the secure two-party computation using Boolean circuits. As a cryptographic privacy-preserving technique, GC protocol enables secure evaluation of a function expressed as a Boolean circuit composed of binary gates.  In this process, the inputs and outputs of each gate are masked, ensuring that the party evaluating the GC cannot access any information about the inputs or intermediate results during the function's evaluation, thereby securing against honest-but-curious adversaries. 

 Songhori \emph{et al.}
\cite{songhori2015tinygarble} design a sequential circuit description tool for generating and optimizing compressed Boolean circuits used in secure computation, such as Yao’s GC  \cite{yao1982protocols}. As shown in \cite{kondi2017privacy}, Boolean formulas can be garbled in a privacy-free setting, where no ciphertexts are produced. To improve computing efficiency, a GC accelerator and compiler are developed in \cite{mo2023haac} to reduce computing overheads in practical privacy-preserving computations. 
GC-based methods demonstrate effectiveness in supporting confidential computing, controlling data usage, and processing arbitrary functions. However, GC-based approaches with affordable bitwise computations for binary operation-oriented applications in power systems are still in early development. Hardware-based methods are less susceptible to certain types of software vulnerabilities (e.g., malware or hacking attacks), making them a valuable complement or alternative to software-based power system applications. Therefore, the integrated design of hardware-software methods for enhanced privacy protection and cybersecurity is a viable future research direction. 

% \newpage
\scriptsize
%\begin{longtable}{m{3em} m{2em} m{6em} m{9.5em} m{18em}} 
\begin{longtable}{ m{2.7em}  m{1.7em}  m{6em} m{5.3em} m{7.8em} m{14.1em}} 
\caption{Miscellaneous and Emerging Scalable and Privacy-Preserving Methods.}
    \label{table_others_literature} \\
    \toprule
Method & Ref. & Problem  & Structure  & Adversaries & Key Features\\
    \midrule
    \endfirsthead
    \caption[]{(Continued)}\\
    \toprule
Method & Ref. & Problem  & Structure  & Adversaries & Key Features\\
    \midrule
    \endhead
    \bottomrule
    \endfoot

\midrule
    SD  & \cite{wang2019privacy} & Average consensus & Distributed &  Honest-but-curious agents, eavesdroppers & 1-Each agent decomposes its state into two substates and only one substate is visible to others; 2-achieve exact consensus.\\

    SD  & \cite{wang2021privacy} & Average consensus & Distributed &  Honest-but-curious agents, eavesdroppers & 1-Each agent is decomposed into a few homologous subagents; 2-the homologous subagents exchange information directly, while the non-homologous subagents communicate via encrypted messages; 3-final group decision value is protected.\\

    SD  & \cite{zhang2022privacy} & Average consensus & Distributed & Byzantine agents$^\star$, malicious agents, eavesdroppers & 1-Time-varying digraph with bounded number of adversarial agents; 2-consider worst-case malicious agents.\\

    SD  & \cite{zhang2022privacy_wang} & Dynamic average consensus & Distributed & Honest-but-curious agents, eavesdroppers & 1-Agents cooperatively track the average of local time-varying reference signals; 2-convergence guaranteed; 3-applications on formation control of mobile robots.\\

  SD &\cite{sun2023privacy} & Distributed economic dispatch & Distributed & Honest-but-curious agents, eavesdroppers, $\epsilon$-DP & 1-SD is carried out at each iterative step; 2-hybrid of SD and addition of Laplacian noise.\\

  SD &\cite{chen2023privacy} & Average consensus & Distributed & Honest-but-curious agents, eavesdroppers & 1-A privacy-preserving push-sum algorithm with communication over directed graphs; 2-new definition of privacy preservation.\\

% SD &\cite{} & & Distributed & Honest-but-curious agents, eavesdroppers & 1-; 2-.\\
    
 \midrule

NI &\cite{mo2016privacy} & Average consensus & Distributed &  Maximum likelihood estimate & 1-Provide exact mean square convergence rate; 2-characterize the covariance matrix of the maximum likelihood estimate.\\

NI &\cite{charalambous2019privacy} & Average consensus & Distributed & Honest-but-curious nodes  & 1-Privacy-preserving ration consensus under time-varying delays; 2-exact average; 3-agent update information states using constant positive weights and adding an offset.\\

NI &\cite{manitara2013privacy} & Average consensus & Distributed & Honest-but-curious agents & 1-Characterize the mean square convergence rate of the consensus; 2-derive the covariance matrix of the maximum likelihood estimate on the initial state.\\

NI & \cite{mao2020privacy} &  Economic dispatch & Distributed & Eavesdropper  & 1-A privacy-preserving distributed optimization algorithm over time-varying directed communication networks; 2-add conditional noises to the exchanged states. \\

NI & \cite{li2020privacy} &  Distributed signal processing & Distributed & Honest-but-curious agents, eavesdropper & 1-Use subspace perturbation for privacy-preserving distributed optimization; 2-insert noise in the non-convergent subspace through the dual variable; 3-preserve accuracy.\\

  % NI &\cite{manitara2013privacy} & Average Consensus & Distributed &  & 1-; 2-.\\

\midrule

 GC &\cite{ben2008fairplaymp} & Secure multi-party computation & (N/A) & Semi-honest adversary (coalition of at most $\lfloor n/2 \rfloor$ corrupt players) & 1-Compile the function into a description as a Boolean circuit; 2-perform a distributed evaluation of the circuit while revealing nothing else but the result of the function.\\

 GC &\cite{songhori2015tinygarble} & Secure multi-party computation & (N/A) & Semi-honest adversary & 1-Generate and optimize compressed Boolean circuits; 2-provide scalable emulations via sequential circuit description.\\

GC &\cite{mo2023haac} &  Privacy-preserving computation  & (N/A) & (N/A) & 1-Propose a GC accelerator and compiler to mitigate performance overheads; 2-hardware-software co-design that expresses arbitrary GCs programs as streams.\\

GC &\cite{heath2020stacked} & Privacy-free garbling scheme  & (N/A) & All probabilistic polynomial time adversaries$^\dagger$ & 1-Improve GC-based zero-knowledge proof statements with conditional clauses; 2-computation cost is linear in the size of the codebase and communication is constant in the number of snippets.\\

% NI & \cite{} &  Multi-agent systems & Distributed &  & 1-; 2-. \\

\bottomrule
\end{longtable}
\vspace{-0.4cm}
\scriptsize{\noindent (N/A): Not applicable  ~~ Ref.: Reference ~~ SD: State decomposition ~~ NI: Noise injection ~~ GC: Garbled circuit ~~ $^\dagger$An adversary runs in probabilistic polynomial time algorithm \cite{katz2007introduction} ~~ $^\star$Refers to active adversaries}
\normalsize

% Table \ref{table_others_literature} presents the summary of emerging techniques of scale and privacy-preserving methods, including \textit{state decomposition}, \textit{garbled circuit}, and \textit{noise injection}.

% \subsubsection{Federated learning}
% A Survey on Privacy in Graph Neural Networks: Attacks, Preservation, and Applications

% {\color{red}
% Collaborative learning 

% }

% \begin{figure}[htbp]
%  \centering
%  \includegraphics[width=0.5\textwidth]{example-image-b}
% \caption{Privacy and security issues.}
% \label{fig}
% \end{figure}

% \subsubsection{Quantum computing privacy} 

% $^\star$\footnotesize{Decryption} $^\ast$\footnotesize{Share Generation} 

\section{Future Directions on Scalable and Privacy-Preserving DER Control}
\label{Future_Directions}

With the increasing penetration of DERs, advanced scalable multi-agent control, optimization, and learning frameworks have been developed to adapt to  DER-rich power systems. 
These advancements, driven by the process of DER data across various fields, further increase the power system's vulnerability to privacy breaches and security concerns. 
In this section, we extrapolate new approaches for future scalable,  privacy-aware, and cybersecure pathways to unlock the full potential of DERs, as well as controlling, optimizing, and learning generic multi-agent cyber-physical systems.

% \newpage 

\subsection{Improving Accuracy, Privacy, and Algorithm Efficiency}

Enhancing accuracy, privacy, security, and the efficiency of computing and communication is a key research priority in the design of scalable and  privacy-preserving multi-agent frameworks.
Admittedly, scalability can be achieved via distributed and decentralized  structures that enable parallel computing and communications across agents. However, the local computing costs and agent-to-agent or agent-to-coordinator communications can still be high to pose algorithm efficiency challenges.
For example, in distributed settings, it is crucial to explore accelerated algorithm convergence with reduced communications, such as when each agent interacts with only a limited number of its neighbors, while in decentralized structures, agents should minimize dependence on the coordinator to efficiently manage resource constraints, especially in situations involving node failures, network partitions, or malicious attacks.
These challenges intensify when controlling DERs in large-scale power systems. Moreover, the computing and communication burdens are further aggravated when integrating extra privacy preservation measures into the algorithm design. 

For example, DP-based methods quantify privacy risks using a rigorous mathematical framework, but they inevitably suffer from the loss of accuracy due to the added noise. Research efforts have been made to limit or eliminate the privacy-accuracy trade-offs for DP-based approaches. Nozari \emph{et al.} \cite{nozari2016differentially}  develop a DP-based distributed functional perturbation framework that bounds the error between the perturbed and true optimizers. This methodology permits the utilization of any distributed algorithm to solve optimization problems on noisy functions while protecting agents' private objective functions.  In \cite{wang2022differentially}, a DP-based distributed stochastic approximation-type algorithm is designed to preserve privacy in solving stochastic aggregative games. Mini-batch methods are used to decrease the influence of added privacy noise on the algorithm’s performance and improve the convergence rate.

In contrast to DP, ED-based methods can attain higher precision at the cost of extra computing loads and increased data volume for communication. This is because ED-based strategies often need to transform real numbers into integers and then compute on large integers with large key sizes, e.g., 1024-bit key size in Paillier's key generation. The intensive mathematical calculations on the large ciphertexts (i.e., encrypting and decrypting data) can also result in communication latency. Compared to ED-based techniques, SS-based methods simplify key management by allowing participants to only manage shares rather than a complex set of keys. SS-based schemes primarily rely on polynomial interpolation and simple arithmetic operations over finite fields, which is less computationally expensive. However, SS-based methods can demand more frequent communications when exchanging shares, especially for multi-agent frameworks. Apart from software-based methods, hardware-based strategies such as GC are also viable in achieving privacy-preserving data analysis, private information retrieval, and secure multi-party computation. Despite efforts made to mitigate computing overhead, GC’s outlook still needs further exploration considering other factors, e.g., low usability and scalability in regenerating circuits. Other emerging obfuscation tools such as NI, are up-and-coming to lift the privacy-accuracy trade-off. To summarize, while there has been great enthusiasm toward balancing or eliminating the trade-offs between accuracy, privacy, security, computing and communication efficiency, developing scalable and privacy-preserving algorithms with comprehensively enhanced performance still requires future efforts.

\subsection{Establishing Trustworthiness Across Fields}

The integration of DERs is creating profound impacts on the electric power sector, fostering a highly interconnected community with \emph{everything as a grid} \cite{NREL_electrification}.
% 
% \revfirst{{\bf{REV 1.10}}}{\color{blue}
The highly connected nature of modern power grids requires the establishment of strengthened `trustworthiness' across different fields. 
To this end, the definition of \emph{trustworthiness across fields} can include three key aspects: 1) trustworthy data and features for the artificial intelligence (AI) model; 2) trustworthy analytical results for the DER control; and 3) trustworthy human-machine interaction for the system management.
% 
% }

% {\color{blue}
\emph{Trustworthy data and features for the AI model}. 
As power grids transition alongside the rapid progress of AI, the broad capabilities of AI offer new possibilities for consolidating grid-edge DERs to enhance grid sustainability and resiliency. 
However, the need to collect, process, and transfer sensitive system and customer data for fine-tuning learning models can raise new technical, economic, and ethical risks that have not been seen before. 
To manage risks, the \emph{U.S. National Institute of Standards and Technology} has developed a comprehensive AI risk management framework \cite{NIST_Risk_Management} related to individuals, organizations, and society. 
Therefore, it is essential to securely collect accurate, unbiased, and representative real-world data for AI models to make fair decisions without violating the privacy and security of the system. 
\emph{Trustworthy analytical results for the DER control}. 
The privacy and cybersecurity challenges in the AI field are propagating into the power energy field, e.g., privacy challenges in natural language processing based on machine learning in the electric energy sector  \cite{majumder2024exploring} and power system fault diagnosis within quantum computing field \cite{fei2024power}. 
% 
% }
Majumder \emph{et al.}  \cite{majumder2024exploring} point out that privacy and cybersecurity emerge as a paramount concern when integrating large language models (LLMs) into electric energy systems. Besides, emerging PGD-based adversarial attacks can cause devastating attack results for LLMs \cite{geisler2024attacking}, resulting in catastrophic failures if deployed in power systems. Moreover, by leveraging the principles of quantum mechanics, quantum computing can break widely-used cryptographic systems by making it possible to factor large numbers efficiently. Zhou and Zhang in \cite{zhou2022noise} show the potential of quantum machine learning in providing resilient and secure decision-making of large-scale power systems. The quantum-inspired methods can enhance security via quantum key distribution \cite{yan2022quantum}, resist quantum computing attacks, and open new possibilities for data transfer and information processing, e.g., quantum cryptography \cite{pirandola2020advances} and quantum communication \cite{gisin2007quantum}.
Therefore, there is a need to test existing frameworks and develop new privacy-preserving and cybersecure solutions by incorporating trustworthy analytical results from various fields to benefit the power and energy community.
% 
% \revfirst{{\bf{REV 1.10}}}{\color{blue}
% 
\emph{Trustworthy human-machine interaction for the system management.}
Enhancing the safety and security of human-machine interaction is also critical for making decisions to manage the DER-rich power systems. 
It requires consideration of  coupled cyber-physical power system architecture \cite{alvarez2024cyber}, the interconnected industrial networks \cite{huang2024toward}, and the stochastic human factors on DER control, such as from the behavioral science perspective.

\subsection{Developing Zero-Trust Standards}
\label{Sec_Zero_Trust}

With the growing requirements on data confidentiality and system integrity, holistic privacy-aware and cybersecure frameworks that can handle both passive and active adversaries are emerging. Importantly, the development of privacy-aware and cybersecure multi-agent frameworks needs to be compatible with various access control, communication, computation, detection, and mitigation techniques.

Toward this goal, we examine the concept of \emph{zero-trust (ZT)} to show the efforts on aligning high-caliber privacy and security standards. ZT is initially proposed to protect resources under the assumption that trust is never implicitly granted \cite{stafford2020zero}. Within ZT, the range of cybersecurity paradigms shifts from static network-based perimeters to a focus on users, assets, and resources.
Moreover, ZT can consolidate a set of guiding principles for workflow, system design, and operations to improve the security posture to any sensitivity level.  The core ZT logical components include 1) a \emph{policy engine} that makes and logs the decision on granting, denying, or revoking access to the device/user, 2) a \emph{policy administrator} that executes the decision from the policy engine and manages the operation of subject-resource communication pathways, and 3) a \emph{policy enforcement point} that communicates with the policy administrator to enforce the enabling, monitoring, and terminating of sessions between a subject and an enterprise resource. As a generic network security model, ZT architecture secures a system’s overall information security, including applications such as cyber supply chain security \cite{do2021integrating}, secure cloud computing \cite{sarkar2022security}, and the industry internet of things \cite{li2022future}.

The increasing adoption of DERs and the ever-complicating adversarial landscape in power systems highlight the need for a holistic privacy-aware and cybersecure framework. Ultimately, it should provide multi-layer internal, external, and hierarchical protection against existing and unforeseen passive and active adversaries, even in the failure of multiple agents or leaders (e.g., the SO, coordinators, and aggregators).
Based on the definition of ZT, \emph{zero-trust architectures (ZTAs)} enforce stringent security measures, where neither local resources nor user identities are automatically trusted solely based on their physical or cyber location.

Research efforts have identified the possibility of deploying ZTAs to manage grid-tied resources in various locations, such as commercial, residential, or governmental areas \cite{he2022survey,alagappan2022augmenting,zanasi2022zero,li2023zero}. In \cite{alagappan2022augmenting}, ZTA is applied to virtual power plants to achieve enhanced protection of virtual power devices.  Zanasi \emph{et al.} in 
\cite{zanasi2022zero} explore the application of ZTA in industrial systems to minimize cyber risks. In \cite{li2023zero}, ZT is applied to enforce identity and access management, securing data communication between EV chargers and cloud platforms while avoiding user-level privacy leakage.  
Despite the established fundamentals of \emph{ZT}, the deployment of ZTAs in large-scale DER control problems is still in its early stages. The challenges include the costs associated with upgrading legacy power system infrastructure, interoperability issues due to varying protocols and standards, potential communication latency, and the significant investment required. In the future, leveraging high-standard privacy and security concepts to develop privacy-aware and cybersecure frameworks that are deployable for power systems will be a challenging research focus.

% \subsection{Quantum privacy}
% talke about the 
% Qutum privacy 

% communications perspective 
% combined with 

% Review progress on 

% privacy issues, cyber-physical security, 

% \begin{table*}[!htb]    \caption{Comparison of PP Approaches on the Basis of Method, Merits, Weaknesses, and Computing Complexity.}
%     \label{table:3}
%     \sisetup{table-format=2.4}% table-column-width =2.5cm
%     \centering
%     \small
%     \begin{tabular} {l S S } 
%         \toprule
% \thead {Method}  
%     & {\thead{dsaasd add dada (sec) \\ 1000 dsad} } 
%         & {\thead{adsdas das  Time (sec) \\ 2000 fdsfds}} \\
%     \midrule
% dfg gdfs & 22.222 &  4.4222 \\
% sdfgg    & 22.222 &  4.0222 \\
% sdfggsd  & 22.222 & 22.2226 \\
% dsfggfd  & 32.355 &  2.277  \\
% dfd      & 20.254 &  2.238  \\
%     \bottomrule
%     \end{tabular}
% \end{table*}

% \input{Section6_ZeroTrust}
\section{Conclusion}
\label{Conclusion}

With the increasing integration of distributed energy resources (DERs)  in large-scale power grids, many power system control, optimization, and learning problems require scalable solutions within a multi-agent framework. Besides, the frequent and mandated exchange of sensitive information among agents makes the entire multi-agent system vulnerable to privacy breaches. These privacy breaches can cause privacy and cybersecurity risks to threaten the function of the entire power grid. Therefore, it is crucial to protect privacy and achieve scalability when deploying multi-agent frameworks for DER control, targeting for greater sustainability, security, and resilience.

This paper provides a comprehensive review of recent advancements in scalable and privacy-preserving multi-agent frameworks from multi-disciplinary research areas, highlighting their applications for controlling DER in power systems. It offers a systematic summary of multi-agent frameworks based on their scalable computing and information exchange structures, 
% The review summarizes and categorizes state-of-the-art scalable algorithms for solving multi-agent problems, 
illustrating their applications in DER control problems across different disciplines. 
This review identifies internal, external, and hierarchical types of adversaries in multi-agent-based DER control problems, including \textit{external eavesdroppers}, \textit{honest-but-curious agents}, and \textit{system operators and/or coordinators/aggregators}. 
Regarding privacy protection, this paper further explores mainstream privacy preservation techniques, such as \textit{differential privacy}, \textit{encryption-decryption-based cryptosystem}, and \textit{Shamir's secret sharing}, along with other and emerging methods such as \textit{state decomposition}, \textit{noise injection}, and \textit{garbled circuits}. 
Recent advancements underscore the significant scalability and privacy preservation capabilities of these approaches for the electric power sector. 
% , and these techniques can be further enhanced along with interdisciplinary techniques and industrial standards and protocols. 
% 
Last but not least, this paper discusses three potential research directions on \textit{improving accuracy, privacy, and algorithm efficiency}, \textit{establishing trustworthiness across fields}, and \textit{developing zero-trust standards}. 
% 

% for large-scale DER-populated power grids.

% In this way, researchers and engineers can catalyze the development and application of multi-agent frameworks for DER control in power grids.

% for privacy-preserving multi-agent frameworks in the realm of DER control

% We believe that this systematic review is a valuable contribution to the adoption and development of scalable multi-agent frameworks with enhanced privacy and cybersecurity for large-scale DER-populated power grids.

\section*{Acknowledgment}

The authors would like to acknowledge the
National Science Foundation under Grant 2220347 and Grant 2145408, the US Department of Energy under award DE-CR0000018 and award DE-EE0009658, and grants from Princeton University's School of Engineering and Applied Science and Andlinger Center for Energy and the Environment, for their support of this work.

\bibliographystyle{elsarticle-num} 
\bibliography{bibliography}




% %% For numbered reference style

%   \textit{\LaTeX: a document preparation system},
%   Addison Wesley, Massachusetts,

% \end{thebibliography}

\end{document}